\documentclass[fleqn,usenatbib]{mnras}

\usepackage{newtxtext,newtxmath}

\usepackage[T1]{fontenc}

\DeclareRobustCommand{\VAN}[3]{#2}
\let\VANthebibliography\thebibliography
\def\thebibliography{\DeclareRobustCommand{\VAN}[3]{##3}\VANthebibliography}

\usepackage{siunitx}
\usepackage{graphicx}	
\usepackage{amsmath}	
\usepackage{subcaption}
\usepackage{soul}
\usepackage[normalem]{ulem} 

\title[TeV afterglow of structured GRB jets]{TeV afterglow emission from a structured GRB jet using the kinetic approach}

\author[J. P. Hope et al.]{
John P. Hope$^{1}$,\thanks{E-mail: jph58@bath.ac.uk}
Hendrik J. van Eerten$^{1}$,
Sayan Kundu$^{1}$,
and Patricia Schady$^{1}$
\\
$^{1}$Department of Physics, University of Bath, Claverton Down, Bath, BA2 7AY, UK\\
}

\date{Accepted XXX. Received YYY; in original form ZZZ}

\pubyear{2023}

\begin{document}
\label{firstpage}
\pagerange{\pageref{firstpage}--\pageref{lastpage}}
\maketitle

\begin{abstract}
Recent years have seen a growing sample of TeV emission detections in gamma-ray burst afterglows, as well as an increasing role for structured jets in afterglow modelling. Using a kinetic approach, we show that the structure of an afterglow jet impacts its TeV emission, with jets where the energy falls off more sharply with angle showing a decrease in Inverse Compton (IC) peak flux relative to synchrotron peak flux. We use a modified version of the code \textsc{katu}, to which we have added adiabatic expansion and a fully self-consistent treatment of IC cooling both for the electron and photon populations. We compare our results to the semi-analytical code \textsc{afterglowpy}, finding a good agreement with our model except at early times off axis where the effects of baryon loading are important. We compare electron cooling in the cases where there is no IC cooling, Thomson cooling and an inclusion of Klein-Nishina effects, finding that the spectra can only be distinguished if the Compton potential is significantly increased. We obtain a similar cooling rate compared to semi-analytical solutions, with some small difference at the transition from IC to synchrotron dominated cooling. Finally, we use best-fit parameters determined by \textsc{afterglowpy} and re-scaled for our model to reproduce the light curves of GRB 170817A. For our choice of parameters, we find that GRB 170817 would not have been detected in the TeV domain if seen on-axis, even by the upcoming Cherenkov Telescope Array Observatory.
\end{abstract}




\section{Introduction}
\label{sec:Intro}
In recent years, gamma-ray burst (GRB) afterglow research has been forced to grapple with two problems: the role of structured jets in observed emission (see e.g. \citealt{StructJetReview} for a recent overview), and the origin of very high energy (VHE) emission detected in several GRBs (as recently reviewed by e.g. \citealt{TeVReviewObs, TeVReviewTh}).

Structured jets with an angle-dependent energy distribution were first discussed in the late 90s/early 2000s, soon after the first discovery of GRB afterglows in 1997 \citep{costa1997discovery, van1997transient}, as a means to unify and/or constrain the different afterglows and energetics that were being observed (see for example \citealt{Mészáros_Struct, Rossi2002, Granot_2003, Zhang_2004}). However, interest waned due to the lack of distinct observational signatures which could be associated to any particular structured jet \citep{granot2006structure}. This changed in 2017 with the multi-messenger detection of GW170817/GRB 170817A \citep{PhysRevLett.119.161101, Abbott_2017}. It quickly became apparent that a jet with a uniform top-hat energy profile would be insufficient to explain the observed temporal evolution, with the afterglow best explained by a structured jet seen off axis (e.g. \citealt{2017Natur.551...71T, Troja2018, 170817AX-ray, Alexander_2018, Wu_2018, Troja2019}).

Structured jets can refer to any generic energy structure, of which several have been proposed (e.g. Gaussian and power-law distributions of energy with angle, see previous citations, but also for example, \citealt{Mizuta_2009, Duffell_2013, Lazzati_2017, Margutti_2018, Geng_2019, DuqueJet}, which connect jet structure to more recent observations and jet simulations). 

The effect of structure is most apparent when the jet is viewed partially (0 < $\theta_{\rm obs} < \theta_{w}$) or fully ($\theta_{\rm obs} > \theta_{w}$) off axis. This introduces an achromatic effect whereby the peak flux and time of the light curve can shift, leading to a prolonged rising light curve while the tip of the jet comes into view. Chromatic effects have also been demonstrated, such as pair production having an attenuating effect in the prompt emission, which itself is dependent on the energy distribution of the structured jet \citep{10.1093/mnras/stae093}.

The detection of very high energy (VHE) photons from GRB afterglows is a more recent development in the field. The first confirmed TeV detection was GRB 190114C \citep{2019GCN.23701....1M, Ajello_2020}, which had a VHE emission of $0.1-1$ TeV from $1-20$ minutes after detection, at a significance of 50$\sigma$ \citep{2019Natur.575..455M}. The previous year, GRB 180720B was seen by the H.E.S.S observatory at the $100-440$ GeV range around 10-12 hours after the initial prompt emission, with earlier observations not captured due to the emission falling outside the detectable range of the telescopes \citep{2019Natur.575..464A}. Other VHE events of note include GRB 190829A, which was observed by H.E.S.S up to $52$ hours after detection, with the inclusion of a reliable VHE spectrum in that time \citep{doi:10.1126/science.abe8560}, and GRB 201216C, which was observed by MAGIC for up to $2.2$ hours after detection and found to have a sub-TeV spectrum in that period \citep{2020GCN.29075....1B, 20.500.11850/524275}.

More recently, GRB 221009A was detected, and has been called the brightest of all time (or "BOAT") due to its extremely high isotropic energy of $\sim10^{55}$ ergs \citep{lesage2023fermi, an2023insighthxmt, Frederiks_2023}. Along with this, a strong TeV signal was seen by the LHAASO observatory, with photons above 3 TeV being seen from $230-900$s after initial detection, and some photons being detected with energies in excess of 10 TeV \citep{doi:10.1126/sciadv.adj2778}. Furthermore, data from the X-ray light curve has been presented as evidence of a shallow power-law structure \citep{doi:10.1126/sciadv.adi1405}.

Most GRBs that have had a significant TeV detection up till now have been long GRBs, although the short GRB 160821B \citep[a low $\sigma$ VHE GRB, see][]{Acciari_2021}, is a notable exception \citep{2016GCN.19833....1S, 2016GCN.19843....1S}. The most prevalent view is that synchrotron self-Compton (SSC) is responsible for VHE photons, and that it may be common to many GRB afterglows (for recent reviews, see \citealt{TeVReviewTh} and \citealt{TeVReviewObs}). However, it can struggle to fit cases such as GRB 190829A, where a direct synchrotron model fits better, despite violating the synchrotron cut-off limit \citep{doi:10.1126/science.abe8560}.

With the advent of the Cherenkov telescope array observatory (CTAO), which has improved sensitivity at TeV energies over current generation telescopes, more VHE GRBs are expected to be observed in the coming years \citep{2019scta.book.....C}, increasing the need for better modelling of these events.

Modelling the SSC component must account for the up-scattering of photons of frequency $\nu$ (here in the fluid frame) by electrons of energy $\gamma_{e}$, the subsequent energy lost by electrons in this process, and the cross-section of the scattering events. This is further complicated by the Klein-Nishina (KN) effect, where the scattering cross-section is reduced at high photon energies (where $h\nu \gtrsim \gamma_{e}m_{e}c^{2}$). This attenuating effect must be accounted for when looking at VHE emission (e.g. see \citealt{2009ApJ...703..675N, 10.1093/mnras/stv2033, 10.1093/mnras/stac483}).
Analytic models of SSC emission have seen use over the last few decades \citep[see e.g.][]{1994ApJ...432..181M, Petry_2000, Sari_2001}, though they tended to ignore KN effects on the electron population. More recent papers \citep{2009ApJ...703..675N, 10.1093/mnras/stab911, Clement, McCarthy_2024} do account for this, though they approximate the impact through the Compton $Y$-parameter.

Numerical models are well placed to incorporate IC scattering by directly integrating over the Klein-Nishina cross-section and particle populations, without reference to $Y$ \citep{JonesIC, BlumGould}. Such modelling is readily applied to kinetic codes, which have been used both outside (e.g. see \citealt{1997A&A...320...19M, Vurm_2009}), and within GRB research (eg. \citealt{derishev2021numerical, Rudolph_2023}). In particular, one-zone kinetic modelling, with KN effects, has been done in works such as \citet{2009A&A...507..599P}, \citet{2014A&A...564A..77P} and \citet{Fukushima_2017}, and fitting of GRB afterglows with TeV emission has been done by \citet{Derishev_2021} to GRB 190114C, GRB 201216C by \citet{2022ApJ...931..150H} and GRB 221009A by \citet{banerjee2024camelidaeboatobservationsecond}.

Kinetic modelling of GRB 221009A with the inclusion of a multi-zone structured jet has been recently done through the \textsc{ASGARD} code, with synchrotron, SSC and pair annihilation included \citep{Ren_2024}. However, the modelling of IC cooling is also approximated through the use of the $Y$ parameter in a similar fashion to above.

In this paper, we combine the single-zone modelling of general IC scattering, such as in \citet{Derishev_2021}, and the structured jet modelling of \citet{Ren_2024} by modifying a kinetic code to run multi-component structured jets, with the inclusion of fully self-consistent IC scattering and cooling for the electron and photon populations. We used the blazar kinetic code \textsc{Katu} (\citealt{Katu1, Katu2}), due to its readily available library of particle processes and implicit solver. To this, we added support for KN-corrected IC cooling and adiabatic expansion, and then built a shell model into the code to represent the fluid evolution. By running multiple concurrent simulations based off the energy distribution of the jet, we could then obtain a representation of a structured jet, and calculate the total observed flux with respect to the equal arrival time surface (EATS). \textsc{Katu} also has support for hadronic species and particle processes, though we limit this work to looking at electron and photon populations only (examples of leptohadronic modelling of GRB TeV emission can be found in \citealt{2024ApJ...977..242K, banerjee2024camelidaeboatobservationsecond}).

The remainder of this paper is organised as follows: in section~\ref{sec:Method} we outline the kinetic model used and the changes made to it. We also lay out the shell model implemented to calculate the evolution of the fluid, and the approach taken in calculating the observed flux. In section~\ref{subsec:Res_12}, we compare a series of different jets and observer angles with respect to \textsc{afterglowpy}, and discuss the similarities and differences between the two models. In section~\ref{subsec:KNvTM}, we show the resultant TeV light curves, and the impact on IC emission due to jet structure. We then compare the impact of Thomson and KN cooling to the synchrotron-only case, contrasting our numerical approach to the semi-analytical approach taken by others such as \citet{McCarthy_2024}. In section~\ref{subsec:TestCase}, we then fit our model to GRB 170817A through the use of best fit parameters from \textsc{afterglowpy} \citep{afterglowpy2} with re-scaling \citep{vanEerten2012, ScalingRyanEerten}, and obtain a theoretical TeV light curve for 170817A. In section~\ref{sec:Discuss}, we discuss some of the results further and in section~\ref{sec:Conc} we give our concluding remarks.

Throughout the paper, we assume a standard $\Lambda$CDM cosmology with $H_{0} = 67.4$, $\Omega_{m} = 0.315$ and $\Omega_{\Lambda} = 0.685$.

\section{Method}
\label{sec:Method}
We begin by outlining the kinetic model used for this work, and the modifications made to it to allow us to simulate GRB afterglows. We then describe the shell model utilised for the evolution of the fluid, and how we implement the shell and kinetic models into a multi-component jet. Finally, we discuss the approach taken to calculate the observed flux from the resultant photon population, including the effects of being off axis and the attenuation of TeV photons by cosmic microwave background radiation \citep{Stecker_2006}.

\subsection{Kinetic modelling of fluid}
\label{subsec:Katu}
We consider a self-consistent evolution of the electron and photon populations via their respective kinetic equations. 
The electron distribution is evolved by considering adiabatic, synchrotron and SSC processes in addition to the injection of electrons from the forward shock travelling into the upstream circumstellar or interstellar medium. 
The photon population, on the other hand, is evolved via the injection of radiation produced by the electrons due to the above-mentioned energy losses, synchrotron self-absorption of photons by the electrons and the escape of photons from the fluid zone. While the photons are allowed to escape within a certain timescale, it is assumed that the escape timescale of the electrons is significantly larger than the adiabatic timescale of the system, and so is neglected \citep{Katu1}.

To model the evolution of these populations, we use the kinetic code \textsc{Katu}, which was developed to fit leptonic and lepto-hadronic models to blazar spectra \citep{Katu2}. \textsc{Katu} solves a set of kinetic equations, the general form of which is given by
\begin{equation}
    \frac{\partial n_X}{\partial t} = Q_{e} + Q_{i} + \mathcal{L} - \frac{n_X}{\tau_{\rm esc}} - \frac{n_X}{\tau_{\rm dec}} - \frac{\partial}{\partial\gamma_{e}}\left(\skew{-6}\dot{\gamma_{e}}n_X\right),
    \label{eq:KineticEq}
\end{equation}
where $n_X$ is the particle population of a particle species $X$, $Q_{e}$ is the external injection, $Q_{i}$ is the internal injection, $\mathcal{L}$ is the internal losses, $\tau_{\rm esc}$ is the escape timescale, $\tau_{\rm dec}$ is the decay timescale and the last term represents losses due to synchrotron and (prior to the current work) IC cooling\footnote{This treatment of IC cooling applies to electrons rather than photons. IC gains and losses for photons are obtained by determining corresponding source and sink terms. The current work extends the latter approach to electrons.}.

Since we only consider electrons and photons in this work, which do not decay, we may neglect $\tau_{\rm dec}$. For the electron population, we set $Q_{e}$ as a power-law distribution. No external injection of photons is considered, though there will be internal gains and losses which represent the synchrotron emission and the energy gain from IC upscattering.

Both particle populations are evolved considering different processes. These processes affect the particle populations at different timescales, usually controlled by the concerning process, which makes Eq.~\ref{eq:KineticEq} stiff.
Therefore, \textsc{Katu} employs an implicit scheme for evolving the electron population and an exponential integrator to evolve the photon population.
Moreover, since we must account for all escaping photons which form our observed flux, the temporal step for these two schemes is limited by the photon escape timescale \citep{Katu2}.

\subsubsection{Updates to \textsc{Katu}}
\label{subsec:Katu_changes}

When modelling blazar "blobs", \textsc{Katu} ignores adiabatic expansion by considering a fixed volume, and applies a simplified Thomson prescription for IC cooling \citep{Katu1}. However, these simplifications do not apply to our case; the GRB blast wave region will expand over time, and modelling TeV emission requires that we fully account for IC cooling over all regimes. Therefore, two changes were implemented into the library: the addition of adiabatic expansion, and general IC cooling of the electron population. 

Since the volume of the emitting region increases as the jet expands radially outwards, the total number density of the electron population must correspondingly decrease. By considering the number density and energy density of the population, it can be shown (in appendix~\ref{AppAd}) that Eq.~\ref{eq:KineticEq} in the electron case should be modified to include the term
\begin{equation}
    \label{eq:AdExp}
    \frac{\partial n_{e}}{\partial t}\bigg|_{\rm adb} = -\frac{n_{e}}{\tau_{\rm ad}} - \frac{\left(1-\hat{\gamma}\right)}{\tau_{\rm ad}}\frac{\partial}{\partial \gamma_{e}}(n_{e}\gamma_{e}),
\end{equation}
where $\hat{\gamma}$ is the adiabatic exponent and $\tau_{\rm ad}$ is the adiabatic timescale in the fluid frame, which is defined as
\begin{equation}
    \frac{1}{\tau_{\rm ad}} = 3\gamma\frac{\skew{-11}\dot{R_{\rm sh}}}{R_{\rm sh}} -\dot{\gamma}.
\end{equation}
For the purpose of this work, we assume the electron population is relativistic throughout, and thus set $\hat{\gamma} = 4/3$.

For a complete IC treatment of the electron population, one needs to consider the impact of photon energy on the IC scattering cross-section. 
In the original version of the KATU code, the electron energy loss is computed only in the Thomson regime, which is sufficient if the number of scattering encounters remains low. In this case, the cooling can be approximated as
\begin{equation}
\label{eq:Thom_Cool}
    \frac{d\gamma_{\rm ic, T}}{dt} = -\frac{4}{3}c\sigma_{T}\frac{u_{\gamma}}{m_{e}c^{2}}\gamma^{2},
\end{equation}
where $\sigma_{T}$ is the Thomson cross-section and $u_{\gamma}$ is the energy density of the photon field \citep{BoettcherBook}. In this regime, the electron cooling goes as $\gamma^{2}$, allowing for it to be summed with the synchrotron cooling term.

Such an assumption may not stay valid for TeV emission from the GRB jet and can overestimate the cooling.
Therefore, for our work, we adopt the following loss equation for IC cooling, which takes the variation of the scattering cross-section with photon energy into consideration \citep{JonesIC, BlumGould}:
\begin{equation}
\label{eq:IC_gen_cool}
    \frac{d\gamma_{\rm ic}}{dt} = - \int_{\epsilon_{\rm min}}^{\epsilon_{\rm max}}n_{\gamma}(\epsilon)\int_{\epsilon}^{\frac{4\epsilon\gamma^{2}}{1 + 4\epsilon\gamma}}\epsilon_{s}\sigma_{\rm KN}(\gamma, \epsilon, \epsilon_{s})d\epsilon_{s}d\epsilon,
\end{equation}
where $\epsilon$ and $\epsilon_{s}$ is the pre and post-scattered photon energies normalised to the rest mass energy of an electron and $n_{\gamma}(\epsilon)$ is the pre-scattered photon population. $\sigma_{\rm KN}(\gamma, \epsilon, \epsilon_{s})$ represents the full KN cross-section for up-scattering, and is given by
\begin{multline}
    \sigma_{\rm KN}(\gamma, \epsilon, \epsilon_{s}) = \frac{3c\sigma_{T}}{4\gamma^{2}\epsilon}\left[\vphantom{\frac{\epsilon_{s}^{2}}{\gamma(\gamma - \epsilon_{s})}} 2q_{u}\ln q_{u} + (1 + 2q_{u})(1 - q_{u})\right. \\ 
    \left. + \frac{\epsilon_{s}^{2}}{\gamma(\gamma - \epsilon_{s})}\frac{1 - q_{u}}{2}\right]
\end{multline}
where $q_{u} = \frac{\epsilon_{s}}{4\epsilon\gamma(\gamma - \epsilon_{s})}$, for $q_{u} \leq 1 \leq \frac{\epsilon_{s}}{\epsilon},~\epsilon_{s} < \gamma$. In the low energy limit for incoming photons (relative to the electron energy), the general loss term will converge with the Thomson cooling regime.

We describe the numerical implementation of the adiabatic and complete IC cooling into the electron kinetic equation in appendix~\ref{AppKE}.

\subsection{The shell model}
\label{subsec:Shell}

To fully calculate the evolution of the electron population, we need to determine the external injection from the blast wave as well as the evolving \textbf{B}-field which controls the synchrotron emission. Furthermore, we need this model to determine the Doppler boosting, relativistic beaming and time of the resultant flux for a given observer. To do this, we use a shell model, which can cover both the ultra-relativistic and trans-relativistic regimes, and that has already seen widespread use in prior literature to model the GRB afterglow (see e.g. \citealt{10.1093/mnras/stt872, 2013arXiv1309.3869V, HJvERev, afterglowpy}).

The shell model is based on the fireball model (e.g. see \citealt{ReesMeszaros1992, MeszarosRees1993, Meszaros2002}), although the acceleration phase is neglected, as it occurs before the afterglow is detectable. The fireball takes the form of a thin, homogeneous shell, expanding outwards in the burster or lab frame with an ultra-relativistic Lorentz factor, $\gamma >> 1$. The fireball is given a fixed isotropic equivalent energy $E_{\rm iso}$. Initially, $E_{\rm iso}$ is contained within the bulk ejecta mass $M_{\rm ej}$, but as the fireball expands it encounters either the interstellar or circumstellar medium, and begins to sweep up matter, obtaining a mass $M_{\rm sw}$ from the external medium. Since the shell is expanding adiabatically, the total energy must remain fixed. Subsequently, the balance of energy shifts from the ejecta matter to the swept-up matter as time goes on.

The energy of the fireball is thus divided between the ejecta and swept-up matter, such that
\begin{equation}
    \label{eq:E_iso}
    E_{\rm iso} = E_{\rm ej,~iso} + E_{\rm sw,~iso}
\end{equation}
where $E_{\rm ej,~iso}$ and $E_{\rm sw,~iso}$ are the energies of the ejecta and swept-up matter respectively. $E_{\rm ej,~iso}$ is the kinetic energy of the cold ejecta mass,
\begin{equation}
\label{eq:E_ej}
    E_{\rm ej,~iso} = (\gamma - 1) M_{\rm ej}c^{2}.
\end{equation}
Using a trans-relativistic equation-of-state \citep{Mignone_2005}, the energy in the swept-up matter is computed from the shock-jump conditions across the shock front and the assumption of a homogeneous shell and can be shown to be:
\begin{equation}
    \label{eq:E_sw}
    E_{\rm sw,~iso} = \frac{\beta^{2}}{3}(4\gamma^{2} - 1)M_{\rm sw}c^{2}.
\end{equation}
Initially, when $E_{\rm ej,~iso} >> E_{\rm sw,~iso}$, $\gamma$ is constant, and the fireball is in the coasting phase. Once $E_{\rm sw,~iso} \sim E_{\rm ej,~iso}$, the bulk fluid begins to decelerate as the swept-up matter begins to dominate, and the system forgets the initial coasting phase.

\begin{figure*}
    \centering
    \includegraphics{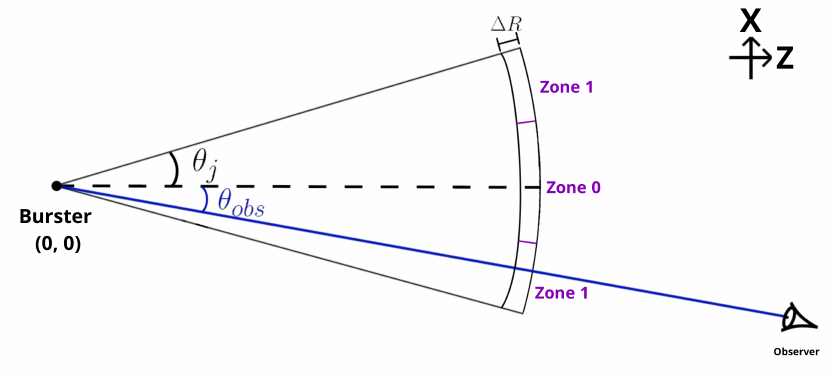}
    \caption{Schematic of a jet with half-opening width $\theta_{w}$ and observer angle relative to jet normal $\theta_{\rm obs}$. The jet is assumed to be moving parallel to $\hat{z}$ without loss of generality. The emitting region of the jet has a width $\Delta R$, and has been split into a circular and annular zone (zones 0 and 1 respectively) to allow for a two component jet to be modelled. More zones ($\sim$18) are required to capture the full structure of a Gaussian or power-law jet.}
    \label{fig:JetSide}
\end{figure*}

The initial Lorentz factor and $M_{\rm ej}$ can influence the time it takes for coasting to end and deceleration to begin. We can take two approaches to determining this. First, we can unilaterally set $\gamma$ across the entire jet and then calculate $M_{\rm ej}$ from Eq.~\ref{eq:E_ej}, accounting for the fact that $E_{\rm iso}$ changes with angle in a non-uniform jet. Alternatively, we can set $M_{\rm ej}$ across the jet as constant, and instead obtain initial Lorentz factors that can vary over angle. We take the former approach for most of this work, though in section~\ref{sec:M_ej_discuss}, we consider the impact of taking the latter approach, which is similar to approaches taken by other works (eg. \citealt{2009ApJ...703..675N, 10.1093/mnras/stab911, McCarthy_2024}).

The shell model can be applied to structured jets by separately evolving the radial flow along different angles, leading to a number of shells representing concentric annuli. To apply the shell model to structured jets, we have applied the generic parameterised models used in \citet{afterglowpy} for the \textsc{afterglowpy} modelling code, which are easy to implement and readily comparable due to their shared parameters and joint normalisation of the core width $\theta_{c}$. The two models used represent a Gaussian and power-law energy profile. For an isotropic equivalent energy $E_{\rm iso} \equiv E(\theta) = 4\pi dE/d\Omega$, their energy distributions are given by
\begin{align}
\label{eq:StructGauss}
    E (\theta) &= E_{0} \exp\left(-\frac{\theta^{2}}{2\theta_{c}^{2}}\right) ~~~~~~~~~~\text{Gaussian} \\
\label{eq:StructPower}
    E (\theta) &= E_{0} \left(1 + \frac{\theta^{2}}{b\theta_{c}^{2}}\right)^{-b/2} ~~~~~\text{Power-law},
\end{align}
where $E_{0}$ is the normalised energy ($E_{\rm iso}$ at $\theta = 0$) and $b$ is the power-law index.

For this shell model, we have neglected jet spreading, as most of our runs occur over times where spreading has minimal impact on the observed flux. The exception to this is when we consider GRB 170817A in section~\ref{subsec:TestCase}, where it accounts for an excess of flux at late times beyond the turnover of the light curves.

The shell width of the forward shock region is determined from mass conservation and shock-jump conditions and given by
\begin{equation}
    \label{eq:Delta_R}
    \Delta R = \frac{R_{\rm sh}}{4(3 - k)\gamma^{2}},
\end{equation}
where $R_{\rm sh}$ is the shock distance from the burster and $k$ is the external density parameter, such that the external density $\rho_{\rm ext} \propto R_{\rm sh}^{-k}$.
Since we are assuming a thin shell, we apply a simple approximation whereby the photons are escaping from a cylindrical region, which then sets the escape timescale for the photons as
\begin{equation}
    \tau_{\rm esc} = \frac{\pi}{4}\frac{\gamma \Delta R}{c},
    \label{eq:tau_esc}
\end{equation}
which also sets the upper bound of our time-step in \textsc{Katu}, and informs the calculation for the observed flux \citep{Katu1}.

It is assumed that the blast wave of the jet is a strong shock due to the pressure and density difference expected between the downstream shell and the cold external medium. Subsequently, we can take the related shock-jump conditions, and assuming the equation of state outlined in \citet{vaneerten2013gammaray}, we can obtain the relations for the mass density $\rho$ and internal energy density $e$:
\begin{align}
    \label{eq:SJ_start}
    \rho &= 4\gamma\rho_{\rm ext}, \\
    \label{eq:SJ_mid}
    e &= 4\gamma(\gamma - 1)\rho_{\rm ext}c^{2}.
\end{align}

The Lorentz factor of the shock $\Gamma_{\rm sh}$ can be related to $\gamma$ by
\begin{equation}
    \label{eq:Gamma_sh}
    \Gamma_{\rm sh}^2 = \frac{(4\gamma^{2} - 1)^{2}}{8\gamma^{2} + 1}.
\end{equation}

The \textbf{B}-field strength in the post-shock fluid is directly informed by Eq.~\ref{eq:SJ_mid}. Assuming that the energy density of the \textbf{B}-field is a fixed fraction $\epsilon_{B}$ of the total internal energy density of the system,
\begin{equation}
\label{eq:B-field}
    \frac{B^{2}}{8\pi} = \epsilon_{B}e.
\end{equation}

The bulk Lorentz factor and shock distance are evolved in the lab frame using a 4th order Runge-Kutta routine. We jointly integrate $\skew{3}\dot{\gamma}$, obtained from the time differentiated form of Eq.~\ref{eq:E_iso}, along with $\skew{-11}\dot{R_{\rm sh}} = c\beta_{\rm sh}$, according to the lab time-step $\Delta t_{\rm lab} =  \gamma\Delta t_{\rm fluid}$, as constrained by $\tau_{\rm esc}$.

\subsection{Injection parameters}
Due to the presence of a (relativistic) shock, a fraction of electrons that are swept-up are energised by the Fermi-I process (see for example, \citealt{Fermi-1, Achterberg2001, Fermi-3}), which generates a non-thermal power-law distribution of electrons (eg. \citealt{Sironi_2011, Sironi_2013}). These are subsequently injected downstream into the post-shock fluid with a number density distribution given by
\begin{equation}
    \label{eqn:InjDist}
    Q_{e} = C\gamma_{e}^{-p},
\end{equation}
where $p$ is the power-law slope and $C$ is a normalisation constant. From such a particle distribution, and considering a fraction ($\epsilon_{e}<1$) of fluid energy to be contained within those non-thermal electrons, implies that the lower limit of the electron Lorentz factors is given by
\begin{equation}
\label{eq:gamma_min}
    \gamma_{\rm min} = \frac{p - 2}{p - 1}\frac{\epsilon_{e}e}{n_{e}m_{e}c^{2}}
\end{equation}
where $\epsilon_{e}$ is the fraction of the internal energy density given to the electrons and $n_{e}$ is the total number density of electrons in the post-shock fluid.

Since synchrotron cooling $\skew{-6}\dot{\gamma_{e}} \propto \gamma_{e}^{-2}$, at sufficiently high energies the shock acceleration timescale becomes equivalent to and exceeds the cooling timescale, and electrons are cooled down rapidly. This is the synchrotron cut-off limit, which constrains the maximum Lorentz factor which an electron can be injected into the shell. We can find it by setting the radiative losses due to synchrotron emission equal to the energy gained over a single gyration during acceleration. This gives us
\begin{equation}
\label{eq:gamma_max}
    \gamma_{\rm max} = \eta\sqrt{\frac{3q_{e}}{\sigma_{T}}}B^{-\frac{1}{2}} \approx 4.65 \times 10^7 \eta \left( \frac{B}{1 \enskip \rm{ Gauss}}\right)^{-\frac{1}{2}},
\end{equation}
where $q_{e}$ is the elementary charge and $\eta$ is an ignorance parameter which can be scaled to include other unknown effects which might affect $\gamma_{\rm max}$.

To normalise the distribution we can integrate over $\gamma_{e}$ between $\gamma_{\rm min}$ and $\gamma_{\rm max}$ in Eq.~\ref{eqn:InjDist} to find $C$. This, however, still requires a knowledge of the total number density of the injected population per unit time, $Q_{e, 0}$. To obtain this, we must consider the total number of particles being swept-up per unit time into the current volume of the shell. This can be determined from the shell model, which allows us to set the normalisation factor as
\begin{equation}
    C = \frac{p - 1}{\gamma_{\rm min}^{1 - p} - \gamma_{\rm max}^{1 - p}}Q_{e, 0},
\end{equation}
where
\begin{equation}
\label{eq:Inj_rate}
    Q_{e, 0} = 4(3 - k)\gamma^{2}n_{\rm ext}\frac{\beta_{\rm sh}c}{R_{\rm sh}},
\end{equation}
with $n_{\rm ext}$ being the external number density and $\beta_{\rm sh} = v_{\rm sh} / c$, which is the dimensionless velocity of the shock wave. See appendix~\ref{AppAd} for a full derivation of $Q_{e, 0}$.

\subsection{Multi-component modelling}
\label{subsec:Zones}

\textsc{Katu} assumes that the fluid is homogeneous, and considers only the number density of the population of particles. This approximation is fine when considering a top-hat jet, where the energy profile is constant with angle. However, if structure is considered, the energy profile is angle dependent, and the population is no longer homogeneous.

To model a structured jet, the simulation is divided into independent zones, each of which are assumed to be homogeneous. The total energy of the bulk fluid for each zone is then set by the structured energy distribution as given in Eqs.~\ref{eq:StructGauss}~\&~\ref{eq:StructPower}. Each zone solves its own shell model and kinetic equations independently to obtain their respective electron and photon populations. 

The number of zones must be set sufficiently high to appropriately reproduce the structure of the jet. We found that 18 zones was a good compromise between structural resolution and computational speed.

Once this is done, each zone can be set to a global initial bulk Lorentz factor $\gamma_{0}$, that through Eq.~\ref{eq:E_ej} sets the initial $M_{\rm ej}$ proportional to the energy structure of the jet. This also sets the radius of the fireball when it enters the coasting phase \citep{Kobayashi_1999}, $R_{c} = \gamma_{0} R_{0}$, where $R_{0}$ is the initial fireball radius. The corresponding lab time at $R_{c}$ is
\begin{equation}
    t_{\rm lab,~0} = \frac{R_{c}}{c\beta_{\rm sh}}.
\end{equation}
During this coasting phase, all zones have the same fluid and lab times. Once deceleration begins, the different zones will start to diverge in $\gamma$ and co-moving time $t_{\rm fluid}$. This is further complicated by the diverging escape timescales (Eq.~\ref{eq:tau_esc}) and thus differing time-steps in the simulation. To help keep track of each zone and facilitate the calculation of the final flux, we impose the condition that the step size in $t_{lab}$ is the same for each zone for a given step. To achieve this, we track the different $\tau_{\rm esc}$ to determine the most restrictive lab time-step, which then informs the fluid time-steps of every other zone, such that no zone ever moves a time-step larger than its local escape timescale.

Finally, the user can specify the number of data outputs needed for their flux analysis. To accommodate this, an additional constraint is applied to the global lab time-step to ensure the desired number of data points is generated.

\subsection{Observed Flux}
\label{subsec:ObFlux}

\begin{figure}
    \centering
    \includegraphics[scale=0.5]{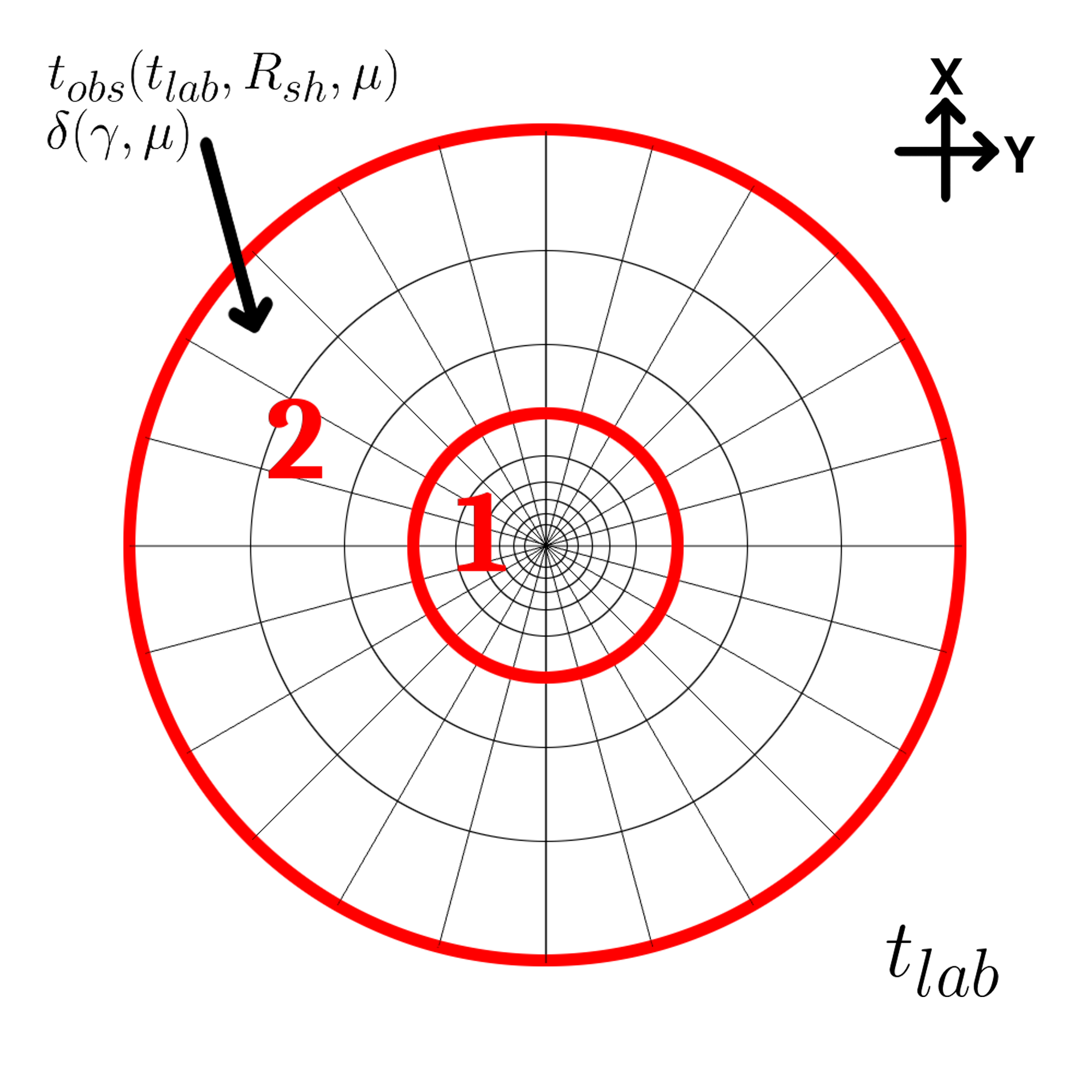}
    \caption{Schematic of the same jet as in Fig.~\ref{fig:JetSide} from the front, at a given lab time $t_{\rm lab}$. At any given time, the front can be subdivided into segments (thin black lines) with a given $(\theta, \phi)$ co-ordinate, with an observer time $t_{\rm obs}$ and Doppler factor $\delta$. The emission coefficient $P_{\nu}$ of a given segment will depend on which zone (solid red lines) it falls into. Each segment may also be subdivided along $\mathbf{\hat{r}}$ to integrate along the radial direction. In this diagram, we show a 2 zone jet, with 216 segments in total. In practise, a large number of zones (18 for structured jets) and around 1000 segments per direction are needed to resolve the flux fully, though if on-axis, the azimuthal dependency can be ignored.}
    \label{fig:JetFront}
\end{figure}

\textsc{Katu} outputs a photon population $n_{\rm ph}(\epsilon_{\rm ph})$ per unit $\epsilon_{\rm{ph}}$, where $\epsilon_{\rm ph} = h\nu / m_{e}c^{2}$ is the normalised photon energy with respect to the electron rest-mass energy\footnote{Here, $\nu$ is in the fluid frame.}. From this, we can determine the power radiated per unit volume per unit frequency as
\begin{equation}
    P_{\nu} = h\frac{n_{\rm ph}\epsilon_{\rm ph}}{\tau_{\rm esc}}.
\end{equation}
We note in passing that the observed flux will not be sensitive to our (approximate) choice of $\tau_{\rm{esc}}$, given that $n_{\rm ph} \propto \tau_{\rm esc}$. 

The observed flux for a given emitting zone is then
\begin{equation}
    \label{eq:FluxGen}
    F_{\nu}(t_{\rm obs}, \nu_{\rm obs}) \approx \frac{1 + z}{4\pi d_{L}^{2}}P_{\nu}\int_{\rm zone} d\Omega dr~r^{2}\delta^{2},
\end{equation}
where $z$ is the cosmological redshift, $d_{L}$ is the luminosity distance, $\Omega$ is the solid angle, $r$ is the distance from origin for the emitting zone and $\delta$ is the Doppler boosting factor (see e.g. \citealt{vanEerten2010, afterglowpy}). While this equation assumes an optically thin medium, in our modelling self-absorption is already accounted for at the local level (modifying $n_{\rm ph}$).

The solid angle accounts for the true extent of the emitting zones of the conical jet (shown in Fig.~\ref{fig:JetSide}). In spherical co-ordinates, where the tip of the jet is aligned with the z-axis without loss of generality,
\begin{equation}
    d\Omega = sin\theta d\theta d\phi.
\end{equation}

Since the origin is set by the burster, the radial co-ordinate will be set as the midpoint of the emitting region $r = R_{\rm sh} - \Delta R / 2$. We can also approximate (since $R_{\rm sh} >> \Delta R$) that $dr \sim \Delta R_{\rm obs}$, the width of the shell in the observer frame, which is given by
\begin{equation}
    \label{eq:delta_R}
    \Delta R_{\rm obs} = \frac{\Delta R}{1 -\beta_{\rm sh}\mu},
\end{equation}
where $\mu$ is the cosine of the angle between an emitting region and the observer, which for a given $(\theta, \phi)$ is
\begin{equation}
    \label{eq:mu}
    \mu = \sin\theta\cos\phi\sin\theta_{\rm obs} + \cos\theta\cos\theta_{\rm obs}.
\end{equation}
Here $\theta_{\rm obs}$ is the angle between the z-axis and the observer. This additional transformation between lab and observer is required to account for the emission time difference between the front and back of the shell, and the direction of the observer relative to the jet normal.

The Doppler boosting factor accounts for the frame transformation between fluid and observer and is given by
\begin{equation}
    \label{eq:delta}
    \delta = \frac{1}{\gamma(1 - \beta \mu)}.
\end{equation}

\begin{figure*}
\centering
\begin{subfigure}{0.48\linewidth}
    \centering
    \includegraphics[scale=0.18, trim={5cm 2cm 0cm 0cm}]{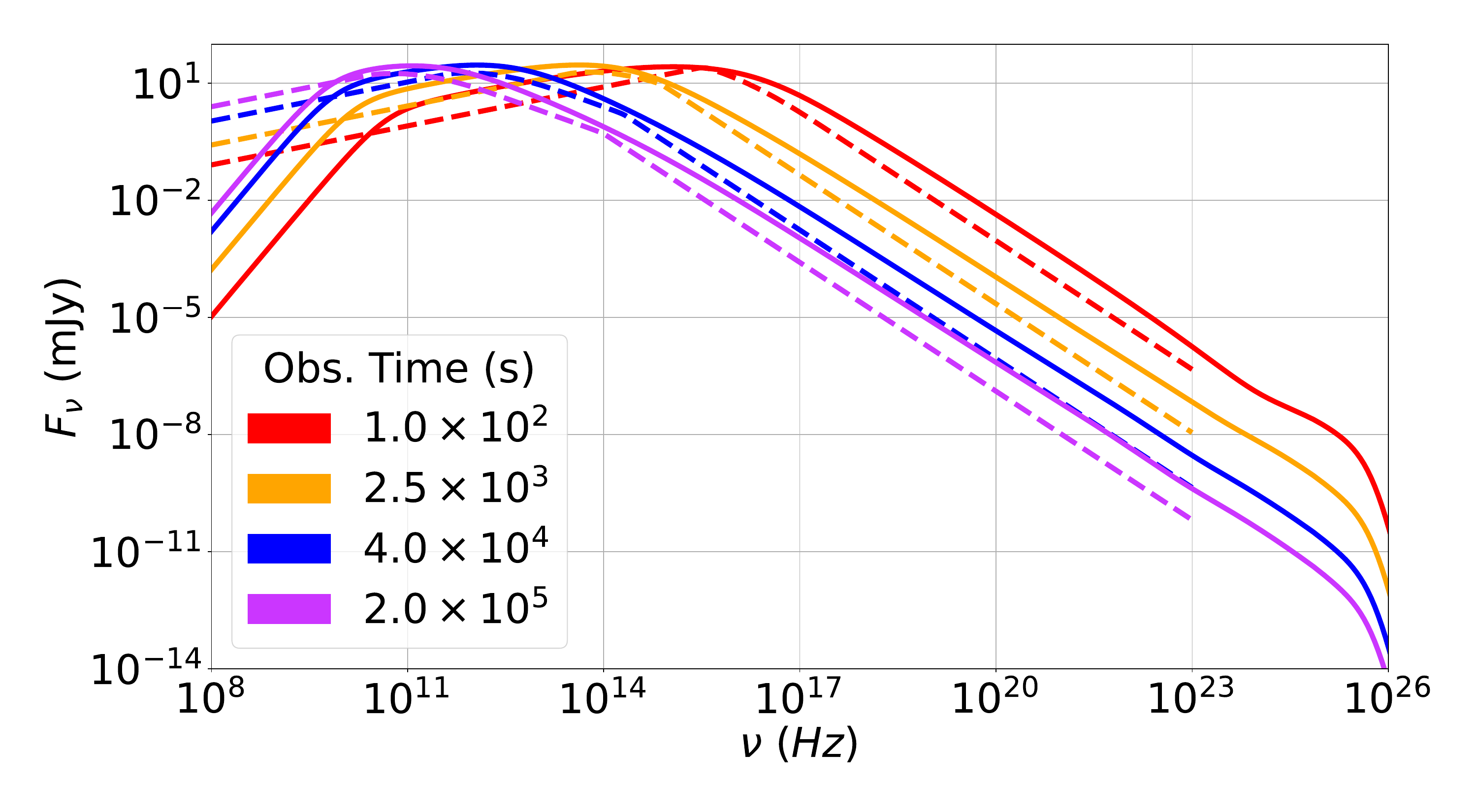}
    \caption{} \label{fig:Spec_A}
\end{subfigure}%
\hspace{0.4cm}
\begin{subfigure}{0.48\linewidth}
    \centering
    \includegraphics[scale=0.18, trim={5cm 2cm 0cm 0cm}]{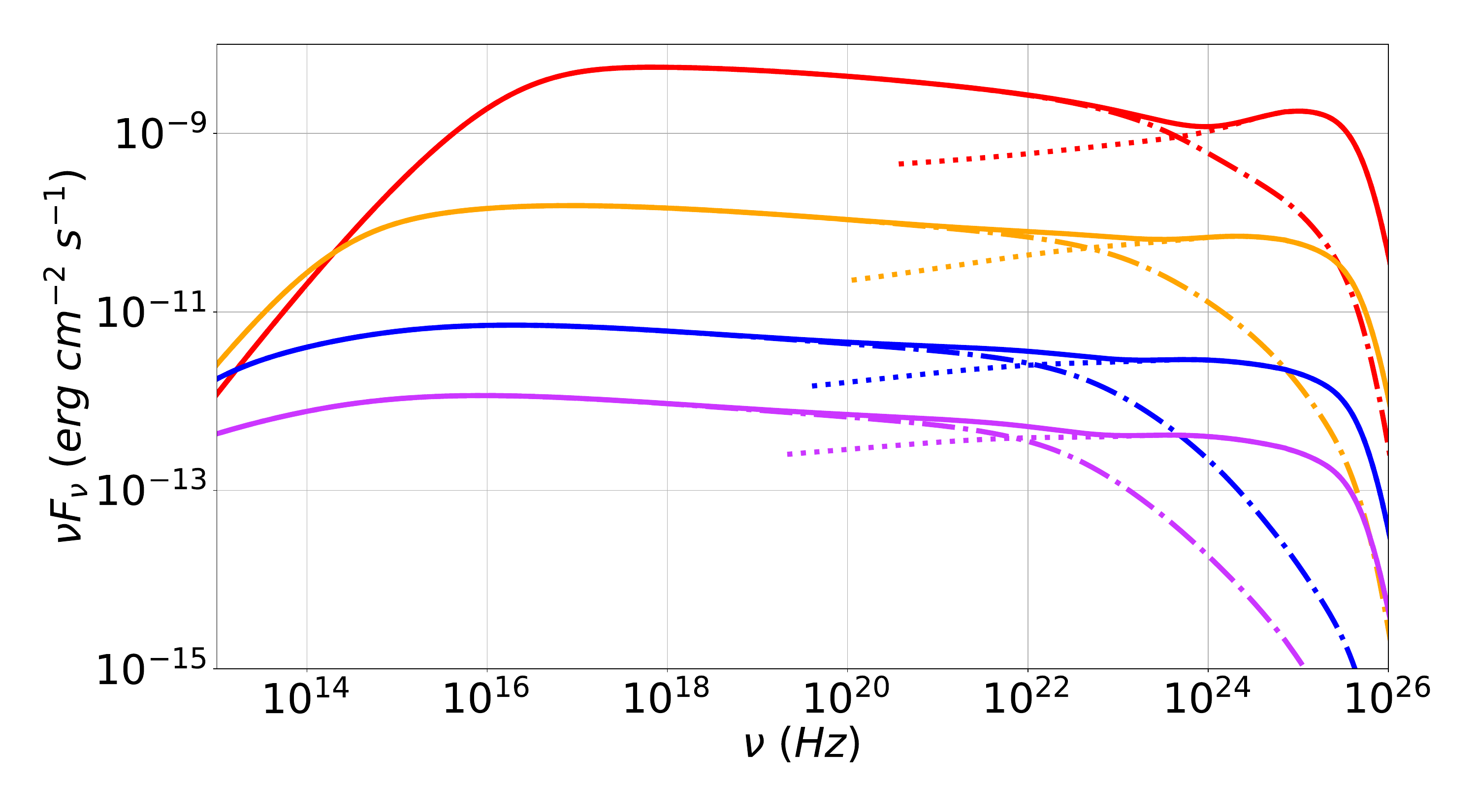} 
    \caption{} \label{fig:Spec_B}
\end{subfigure}%
\caption{Results from the on-axis ($\theta_{\rm obs} = 0$) top-hat run from table~\ref{tab:12_run_params}, for $100$ (red), $2.5\times 10^{3}$ (orange), $4\times 10^{4}$ (blue) and $2 \times 10^{5}$ (purple) seconds. (a) Spectral flux at the given times, from \textsc{Katu} (solid) and \textsc{afterglowpy} (dashed). \textsc{Katu} also includes self-absorption, synchrotron cut-off and IC scattering in its run, while \textsc{afterglowpy} only includes synchrotron cooling. (b) \textsc{Katu} $\nu F_{\nu}$ spectrum of the same run at the same times, with the synchrotron emission (dot-dash), IC emission (dotted) and combined processes (solid).}
\end{figure*}

For a given lab time and zone, the simulation will output $P_{\nu}$. However, photons emitted from different parts of the jet will reach the observer at different times, due to the changing angular position relative to the observer. This is especially important if the observer is off axis to the jet normal since the symmetry about the jet tip will be broken, and $\phi$ becomes important. The surface where the observer time $t_{\rm obs}$ is constant is the equal arrival time surface (EATS), which we want to find for each data point in $t_{\rm obs}$. The observer time for a given $(t_{\rm lab}, r, \mu)$ is
\begin{equation}
    \label{eq:t_obs}
    t_{\rm obs} = \left(1 + z\right)\left(t_{\rm lab} - \frac{r\mu}{c}\right).
\end{equation}

To fully resolve the EATS, we must properly integrate over both its solid angle and radial extent. The latter is especially important for off axis observers, as the EATS may fail to sample flux contributions from the front and back of the emission region if the radial extent is not adequately resolved. To do this, we take the jet, and as in Fig.~\ref{fig:JetFront}, divide it into a number of segments, $I \times J$, with mid-points $(\theta_{i},~\phi_{j})$. A logarithmic grid was used for $\theta$, and a linear grid used for $\phi$. Each segment represents a fixed $\mu$ and, for a given zone, evolving $\gamma$. The radial extent is captured by taking the segment and placing point sources along $\mathbf{\hat{r}}$, such that the $k$th point is given by 
\begin{equation}
\label{eq:radial_integ}
    r_{k} = R_{\rm sh} + \frac{\Delta R}{2K}(1 - 2(k + 1)), ~~ 0 \leq k \leq K.
\end{equation}

We can then integrate over the volume of each segment in a given zone through Eq.~\ref{eq:FluxGen}. Care is taken to ensure no segment overlaps zones in $\theta$ such that $P_{\nu}$ ceases to be constant. The total flux of the segment is also divided among each $r_{k}$ by giving each point $1 / K$ of the flux calculated. Since a given $t_{\rm lab}$ data-set will never align with all $t_{\rm obs}$ points we want to obtain, linear interpolation in log-space is used to find the flux at the desired observer times. Linear interpolation in log-space is also used to translate the flux into the global frequency bins, allowing the total flux across all segments in a zone to be summed at the end.

For an on-axis jet ($\theta_{\rm obs} = 0$), $\mu$ reduces to $\rm cos(\theta)$, allowing $\phi$ to be safely disregarded. In this case, we can set $I = 1500$, $J = 1$ and $K = 10$, to sufficiently capture all flux from the jet. If $\theta_{\rm obs} \neq 0$, then $\phi$ must be considered, which greatly increases the required resolution. This can be partly mitigated by allowing, without loss of generality, $\theta_{\rm obs}$ to be cast along the $x = 0$ plane which, along with the axi-symmetric nature of the jet zones, allows us to omit the lower half of the jet ($\pi < \phi \leq 2\pi$). We instead double the final flux to account for the missing segments. In this case, we found through testing that a good balance between computational expense and resolved flux was obtained by letting $I = 1000$, $J = 500$ and $K = 150$ for the jet edge, and $I = 800$, $J = 400$ and $K = 100$ in the other off axis cases.

Finally, we note that cosmic microwave background radiation can interact with VHE photons through pair production, leading to attenuation by extragalactic background light (EBL) \citep{1992ApJ...390L..49S, doi:10.1126/science.1227160}. This can lead to attenuation of TeV photons for cases beyond $z \gtrsim 0.08$ \citep{Stecker_2006}, and cannot be ignored for GRB afterglows \citep{10.1093/mnras/stad1388}. The effect on the observed flux can be modelled as
\begin{equation}
    F_{\nu,\rm~obs} = F_{\nu,\rm~init} \times e^{-\tau(\nu, z)}
\end{equation}
where $F_{\nu,\rm~init}$ is our flux from equation~\ref{eq:FluxGen} and $\tau(\nu, z)$ is the optical depth for EBL attenuation \citep{doi:10.1146/annurev.astro.39.1.249}. For $\tau(\nu, z)$, we have used the model provided by \citet{10.1111/j.1365-2966.2010.17631.x}, which has then been interpolated to match the redshift and frequency grid in our model.

\section{Comparison with \textsc{afterglowpy}}
\label{subsec:Res_12}

\begin{figure*}
    \centering
    \includegraphics[scale=0.4, trim={1cm 1cm 0 0}]{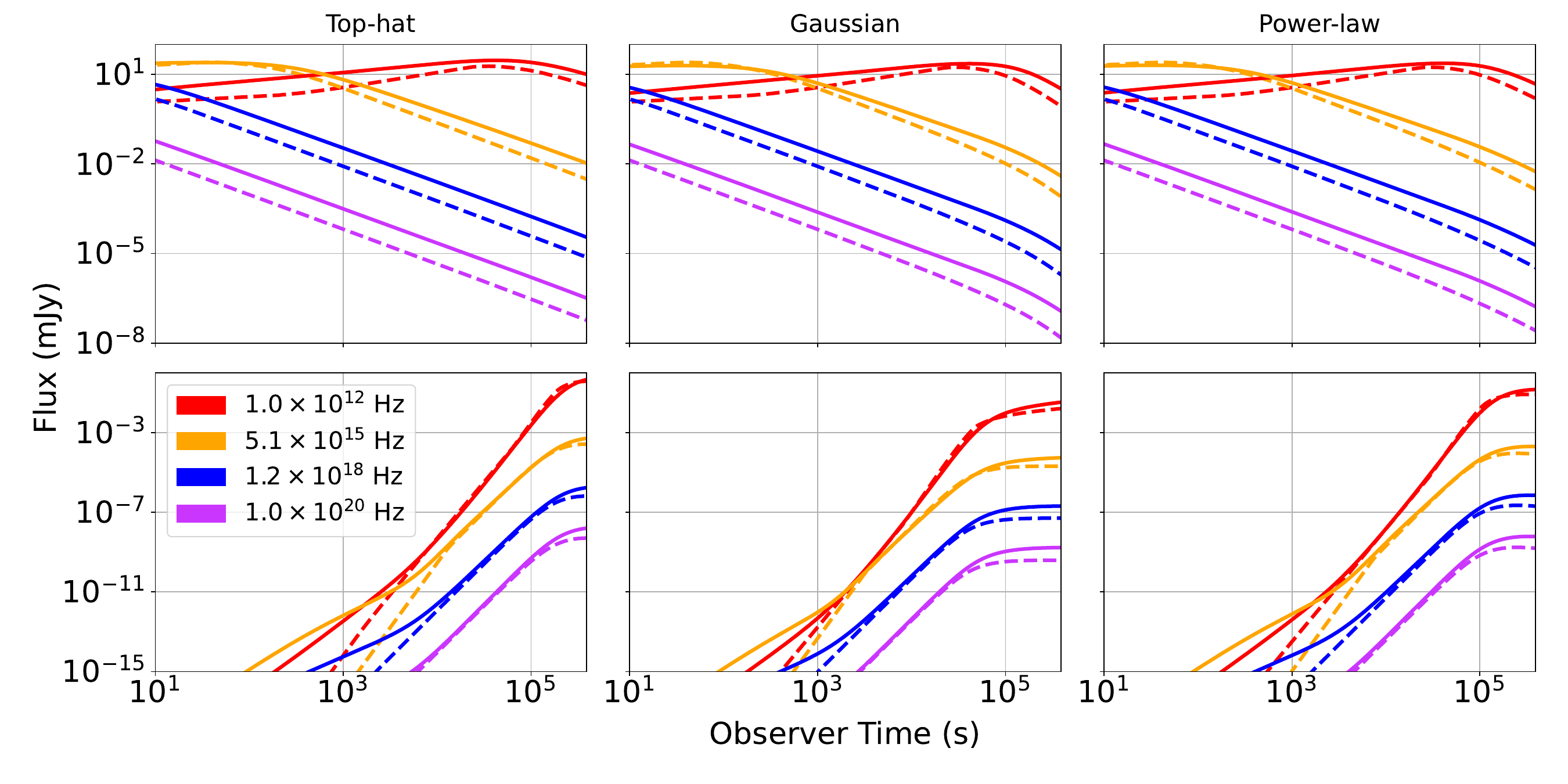}
    \caption{Light curves for the top-hat, Gaussian and power-law jet structures for on-axis (top row) and off axis (0.36 radians or $\theta_{\rm obs} / \theta_{w} = 1.5$; bottom row), from 10 seconds to 4.5 days. The frequencies used are \SI{1e12}{\hertz} (red), \SI{5e15}{\hertz} (orange), \SI{1e18}{\hertz} (blue) and \SI{1e23}{\hertz} (purple). The solid lines are from \textsc{Katu}. The dashed lines are from \textsc{afterglowpy}, using comparable jet types though only including synchrotron emission and no IC cooling. Parameters used can be found in table~\ref{tab:12_run_params}.}
    \label{fig:12_sample_LC}
\end{figure*}

\begin{figure*}
    \centering
    \includegraphics[width=0.48\textwidth]{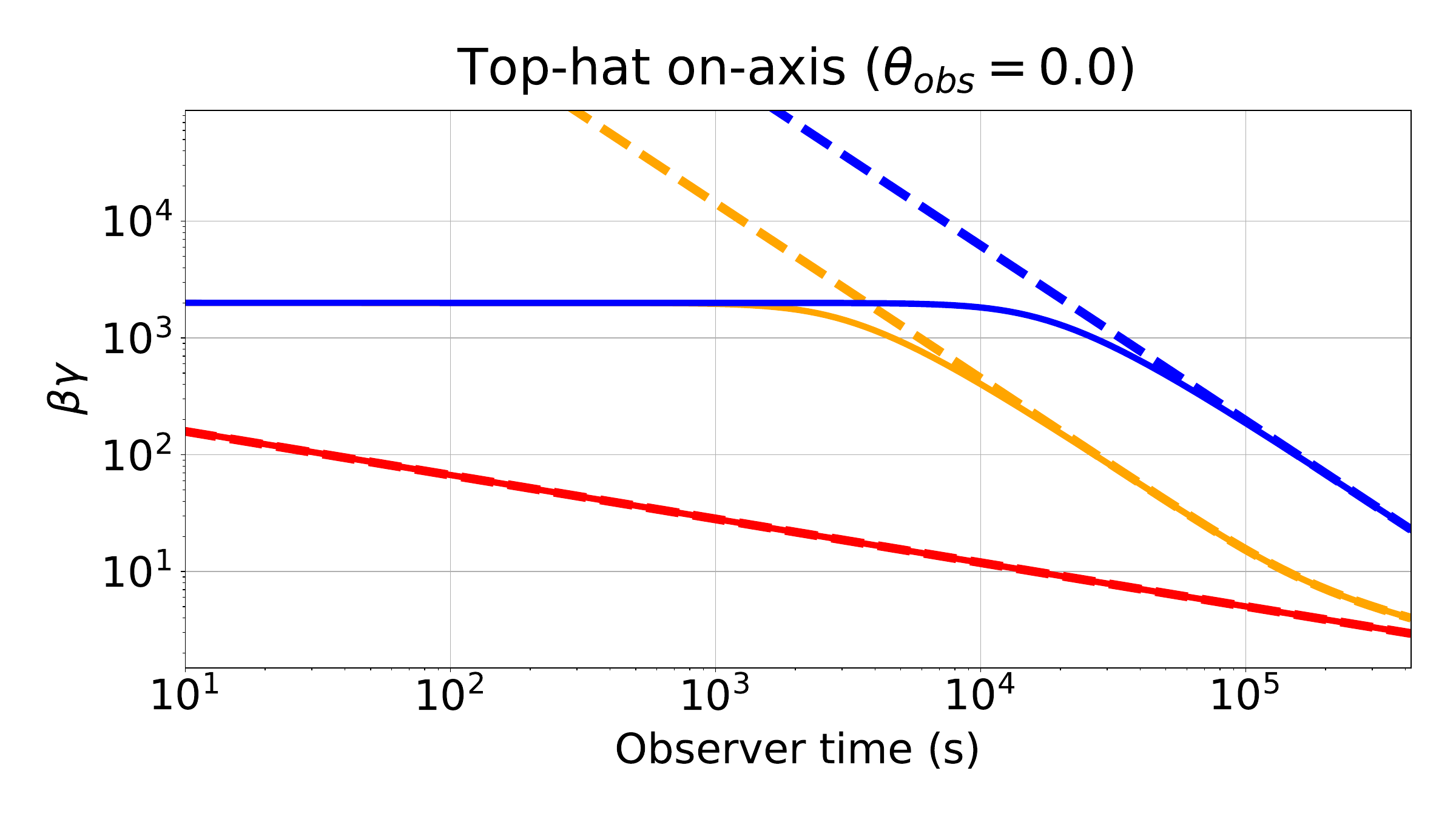}
    \includegraphics[width=0.48\textwidth]{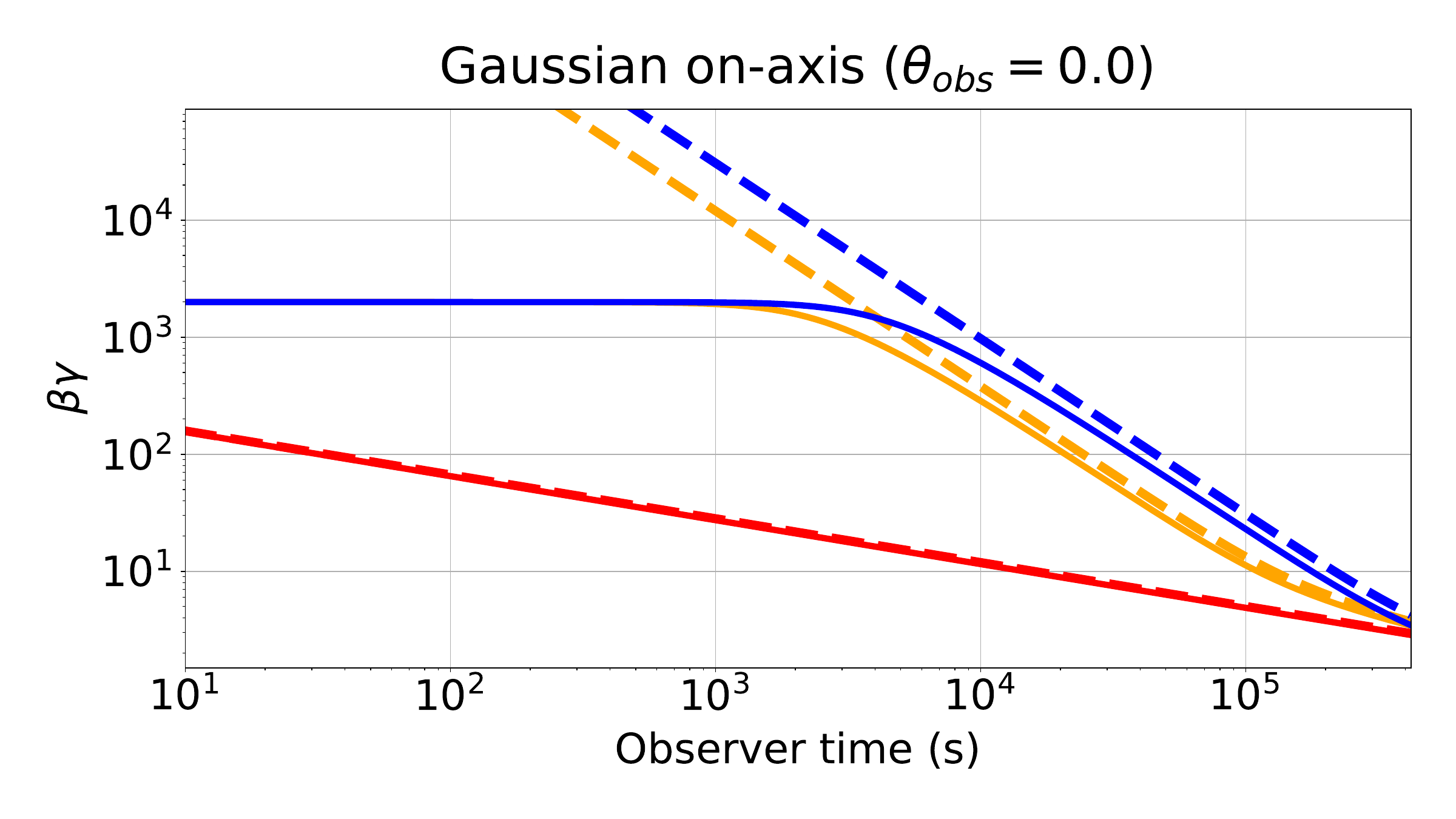}
    \includegraphics[width=0.48\textwidth]{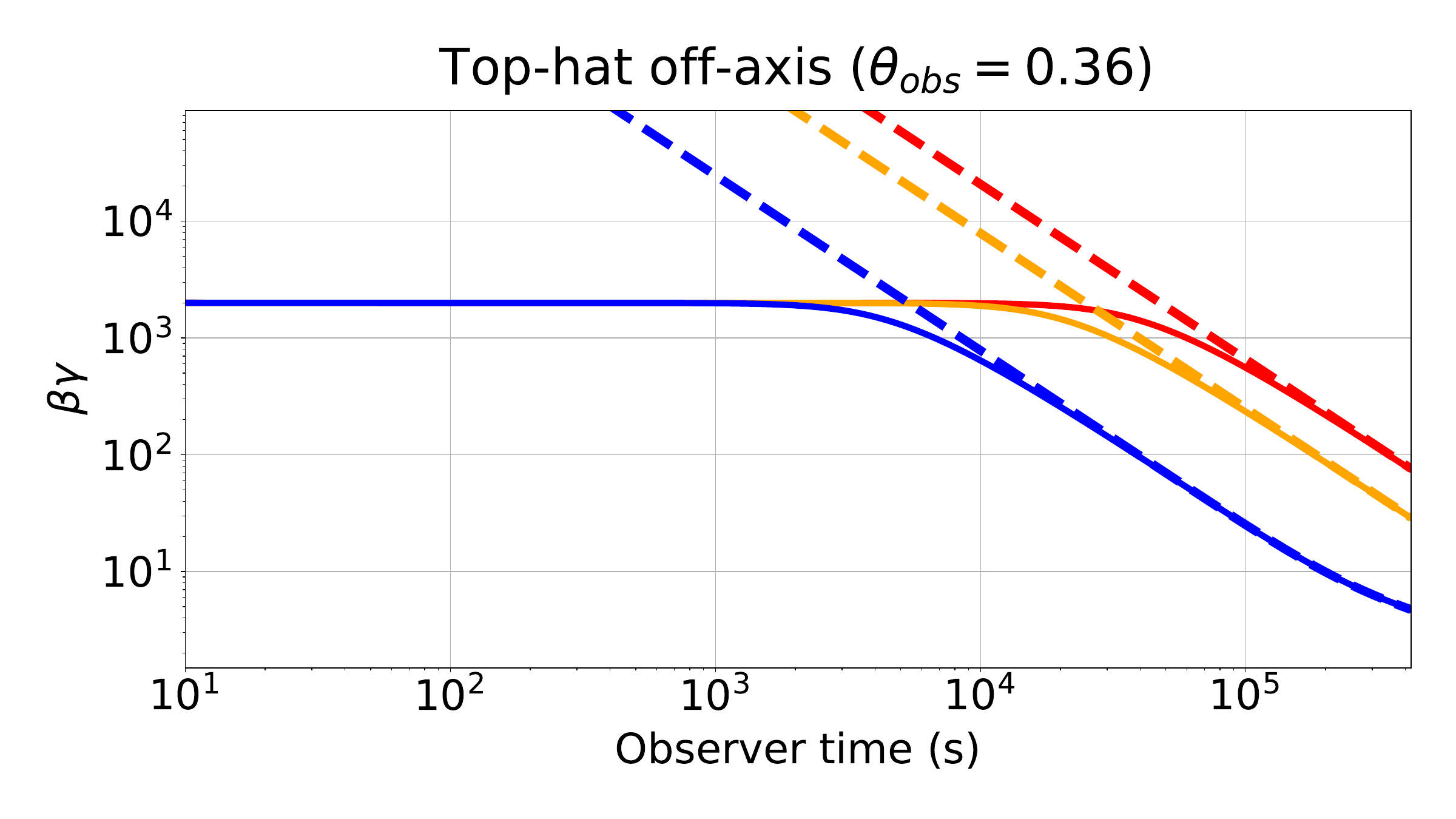}
    \includegraphics[width=0.48\textwidth]{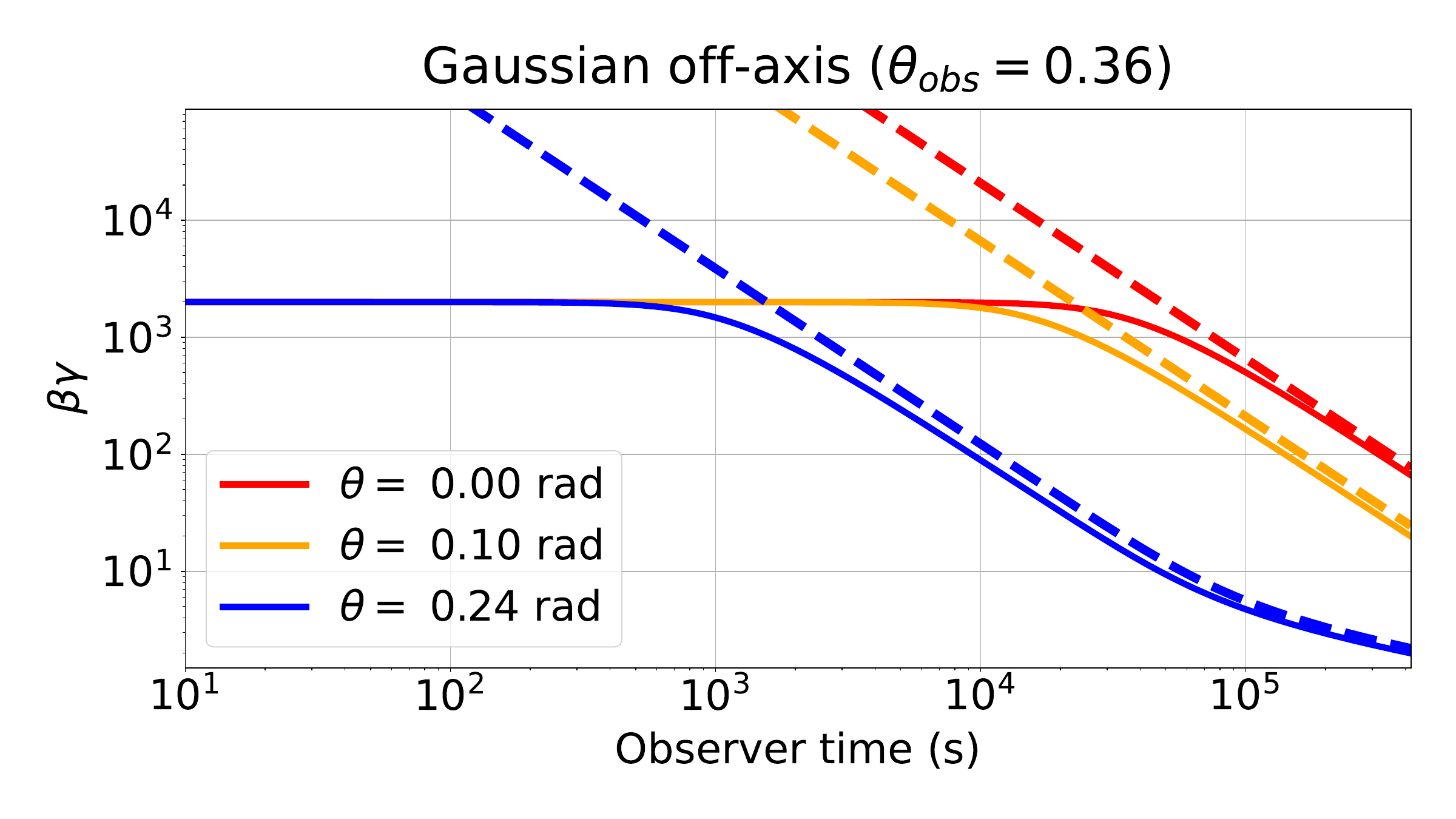}
    
    \caption{Evolution of the shell velocity $\beta\gamma$ in observer time for the top-hat (left) and Gaussian (right) jets from table~\ref{tab:12_run_params}. The top plots are for the on-axis observer, while the bottom plots are for the off axis observer. Solid lines are from the \textsc{Katu} shell model, while dashed lines are from the \textsc{afterglowpy} shell model. We show this for $3$ different positions along the jet, at the tip (red), at $\theta = 0.1$ radians (orange) and at the edge of the jet (blue).}
    \label{fig:shells}
\end{figure*}

We consider a variety of scenarios that allow us to compare different jet structures and observer angles to both \textsc{afterglowpy} and each other. To do this, we examine the 3 different jet structures available in \textsc{afterglowpy}, which are then observed from a selection of angles which go across the jet from on-axis to far off axis: $0$, $0.5\theta_{w}$, $\theta_{w}$ and $1.5\theta_{w}$, making for a total of 12 combinations. The other parameters are kept the same for each run and are provided in table~\ref{tab:12_run_params}. These parameters are selected from \citet{afterglowpy} as an arbitrary set representing a standard long GRB, with the addition of $\gamma_{0}$ and $R_{\rm fire}$ to parameterise the coasting phase in our model. All scenarios were run from $10$ seconds to $4\times 10^{5}$ seconds. In the case of the Gaussian and power-law jets, we set the \textsc{afterglowpy} jet types to include a fixed energy core in the model, as this is closer to the discrete zones used by \textsc{Katu}. We also set the limit of the Top-hat jet to $\theta_{w}$ for the following analysis, with \textsc{afterglowpy} set in a similar fashion for comparison.

For these runs, we have standardised the energy of each jet type for comparison by keeping the jet energy at the tip ($E_{0}$) constant, rather than the total integrated jet energy ($E_{\rm jet}$). This makes it easier to compare the three structured jets together and consider the impact of effects independent of the set parameters. If $E_{\rm jet}$ is set constant instead, it will cause the difference in jet structure to become apparent much earlier, though only in comparison to each other, and will then distract from other effects we wish to consider. From an observational perspective, the actual structure of the jet will only be apparent if the observer is off axis, especially at late times.

\begin{table}
\caption{Parameters used for the comparison runs with \textsc{afterglowpy}, for all structured and off axis jets.}
\label{tab:12_run_params}
\centering
\begin{tabular}{lcc}
\hline
Parameter & Value\\
\hline
$E_{0}$ & $10^{53}$ erg \\
$\theta_{c}$ & $4.58$\textdegree ~ ($0.08$ rad) \\
$\theta_{w}$ & $13.8$\textdegree ~ ($0.24$ rad) \\
$n_{\rm ext}$ & $1$ cm$^{-3}$ \\
$\gamma_{0}$ & $2\times10^{3}$ \\
$R_{\rm fire}$ & $10^{8}$ cm \\
z & $0.5454$ \\
p & $2.2$ \\
b & $2$ \\
$\epsilon_{e}$ & $10^{-1}$ \\
$\epsilon_{B}$ & $10^{-2}$ \\
$\chi_{N}$ & $1$ \\
$\eta$ & $1$ \\
\hline
\end{tabular}
\end{table}

An interesting effect of the parameters used combined with the early start time was that the shocked electrons were injected above the cooling break, where the synchrotron and IC processes dominate over adiabatic cooling (which refers to cooling arising from adiabatic expansion; the cooling term is derived in appendix~\ref{AppAd} as part of the implementation of adiabatic expansion into the code). The runs subsequently start in the fast cooling regime before transitioning to slow cooling over the course of the run. This can be seen in Fig.~\ref{fig:Spec_A}, where a series of spectra over time from \textsc{Katu} (solid lines) are plotted, alongside corresponding spectra from \textsc{afterglowpy} (dashed lines) for the top-hat on-axis case. There is generally good agreement on the expected flux, though \textsc{Katu}'s full treatment of the synchrotron emission leads to some variation in the break frequencies. $\nu_{a}$ is the self-absorption break below which synchrotron photons can be reabsorbed by the electrons and $\nu_{c}$ is the corresponding frequency for the cooling break. However, unlike \textsc{Katu}, \textsc{afterglowpy} does not currently include self-absorption, and thus does not have a break at $\nu_{a}$, leading to the offset in flux before $\nu_{a}$.

$\nu_{c}$, the corresponding frequency for the cooling break can be sensitive to the underlying assumptions used when calculating it. For instance, \textsc{afterglowpy}, along with various other analytical and semi-analytical approaches takes a global cooling approximation where the cooling is assumed to be uniform throughout the shocked material and calculated by equating the lab time with the synchrotron cooling time (eg. \citealt{1998ApJ...497L..17S, vanEerten2010, HJvERev}). Alternatively, local cooling includes information about the varying fluid conditions throughout the shocked region, and has been observed to differ with the values of $\nu_{c}$ compared to the global approach (eg. \citealt{Granot_2002, 2009MNRAS.394.2164V, 10.1093/mnras/stab3509}). \textsc{Katu} takes a hybrid approach; for a given zone, the cooling break is uniform at a given time, but may vary across zones. Some difference between \textsc{Katu} and \textsc{afterglowpy} in the position of $\nu_{c}$ can be expected as \textsc{Katu} includes IC cooling, which will cause a redward shift in the cooling break.

The corresponding spectral break for $\gamma_{\rm min}$, $\nu_{m}$, shows some variation between \textsc{Katu} and \textsc{afterglowpy}, particularly at late times. This difference may seem surprising, as the corresponding injection parameter $\gamma_{m}$ is generally consistent between both codes for relevant time frames. However, the shift can be explained by a difference in the flux calculation: \textsc{Katu} goes a step further by integrating over the radial direction, in addition to $\theta$ and $\phi$ when summing the flux for all shells. This accounts for the difference in arrival times between the front and back of the shell at a given lab time, and is especially important for off axis flux so as to properly sample the entire shell. Removing the radial integration (by setting $r_{k} = R_{\rm sh}$ and $K = 1$ as opposed to Eq.~\ref{eq:radial_integ}) leads to a concurrent $\nu_{m}$ between the two codes. $\nu_{c}$ and $\nu_{a}$ are impacted in a similar way, though have additional issues which can offset them as mentioned above.

The other key difference, explored further in the next section, is the inclusion of IC scattering, which creates the additional bump seen in the spectrum above $10^{24}$ Hz. This can be seen better in Fig.~\ref{fig:Spec_B}, which is the corresponding simulated spectrum in $\nu F_{\nu}$ space. We also show separately the synchrotron (dot-dashed line) and IC emission components (dotted line) that contribute to the total flux. This allows us to see the synchrotron cut-off, which is not included in \textsc{afterglowpy}, and occurs far below the TeV range. We can also see the transition point from the synchrotron to IC regime, starting at $10^{24}$ Hz before moving to $10^{22}$ Hz by the end of the run.

A selection of light curves across the broadband spectrum is shown in Fig.~\ref{fig:12_sample_LC}. Here, the top row is on-axis ($\theta_{\rm obs} = 0\theta_{w}$), while the bottom row is far off axis ($\theta_{\rm obs} = 1.5\theta_{w}$). \textsc{Katu} results (solid) are presented alongside those generated by \textsc{afterglowpy}. For all on-axis light curves, we get a general agreement on the temporal slope and evolution of the light curves, though we consistently see a greater overall flux compared to \textsc{afterglowpy}, especially at higher frequencies. As with the spectra, it is caused by the differing treatment of the spectral breaks, and the systematic offsets that arise from this. The Gaussian and power-law jets show similar light curves to the top-hat due to them all being normalised to the same $E_{0}$, which dominates the on-axis emission. Only at very late times, when the full jet comes into view, would any difference between the three become significant.

When looking at the off axis results, we also see a good agreement with \textsc{afterglowpy} beyond $10^{4}$ seconds. Before this, we observe an excess of flux from \textsc{Katu}. 
Such an excess of flux is caused by the inclusion of an ejecta mass in our shell model, which causes the bulk Lorentz factor to undergo an initial coasting phase before deceleration. We explore the difference between the four-velocities of \textsc{Katu} (solid) and \textsc{afterglowpy} (dashed) in Fig.~\ref{fig:shells}, where we see how different parts of the jet evolve in the observer frame. At early observer times, due to the coasting phase, the fluid Lorentz factor is lower in the $\textsc{Katu}$ model. This reduces the relativistic beaming and allows more flux to be observed at the start. The strength of this effect on the flux output then depends on two factors, $\mu$ and the structure of the jet.

\begin{figure*}
    \includegraphics[scale=0.35, trim={0.5cm 1.5cm 0 0}]{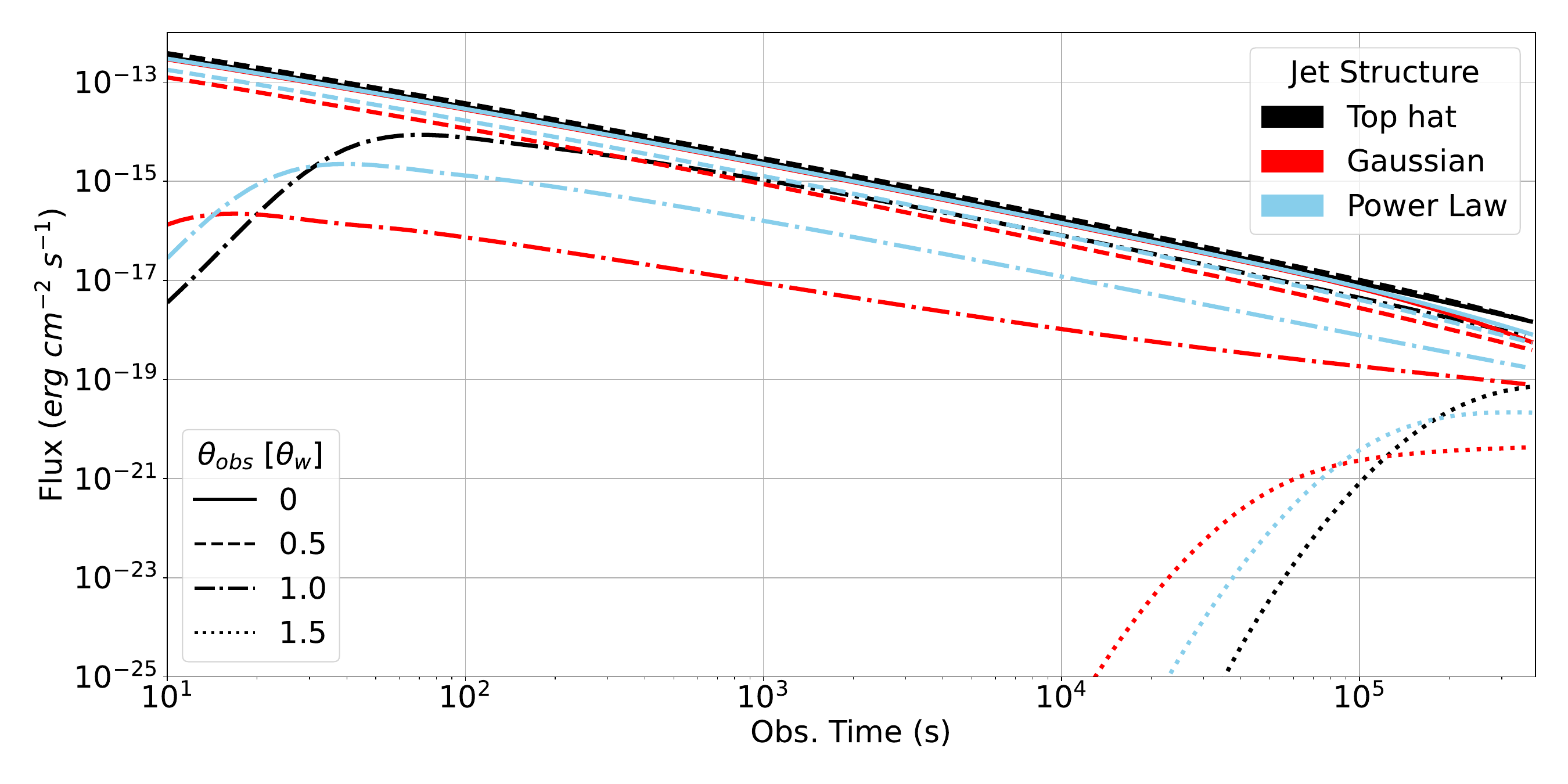}
    \caption{2 TeV light curves for a series of observer angles (0, 0.5, 1 and 1.5 times the jet width $\theta_{w}$) relative to the jet normal for a top-hat (black), Gaussian (red) and power-law (blue) jet structure. Due to the identical use of $E_{0}$ across the runs, the on-axis light curves are practically indistinguishable in the plot. The 0.5 $\theta_{w}$ on-axis light curve is identical to the on-axis top-hat due to the homogeneous energy structure. Parameters used can be found in table~\ref{tab:12_run_params}.}
    \label{fig:LC_2TeV}
\end{figure*}

Starting with $\mu$, we observe that no such excess of flux occurs in the on-axis case, but only arises in the off axis case. While the entire jet decelerates at the same lab time (in the top-hat jet structure), the difference in $\mu$ between the line of sight of the observer and a particular annulus of the jet accounts for the extra time required for emission across the jet to travel to the observer compared to line of sight emission, even if the emission time in the lab frame is the same. This then leads to a disparity from the observer's perspective when each part of the jet is decelerating (ie. Eq.~\ref{eq:t_obs}). When looking on-axis at the jet tip, $\mu \sim 1$, and the jet tip decelerates much earlier in the observer frame, such that \textsc{Katu} and \textsc{afterglowpy} converge before the start of the simulation, and no difference is observed in the flux. If observing off axis, there will be a small decrease in $\mu$ due to the offset between the observer (at $1.5 \theta_{w}$) and the edge of the jet (at $\theta_{w}$), which, when boosted by the $R_{\rm sh} / c$ factor, leads to a large difference in the perceived time of deceleration. In this case, we see that (bottom left Fig.~\ref{fig:shells}) the two shell models converge only at $10^{4}$ seconds, the same time that the flux of \textsc{Katu} and \textsc{afterglowpy} converge.

While the top-hat jet has a constant energy and Lorentz factor structure, the Gaussian and power-law jets do not, adding a further complication to this analysis. As we discussed in section~\ref{subsec:Shell}, when setting the starting conditions, we fixed the initial Lorentz factor, and then by applying the relevant energy structure to the jet, we set $M_{\rm ej}$ by using Eq.~\ref{eq:E_ej}. The effect of this is to cause $M_{\rm ej}$ to also be proportional to $E(\theta)$, and since the external medium is constant at every point along the jet surface, we get different deceleration times in the \textit{lab} frame as well. If the jet is then observed off axis, the edge of the jet will become visible at an earlier time relative to the top-hat case, as the edge of the jet will decelerate earlier in the lab frame due to the lower $M_{\rm ej}$ and thus lower amount of swept-up matter required for the energy in $M_{\rm sw}$ to dominate the shell. This means that the edge will be seen first for a fixed $\mu$ at an earlier observer time (bottom-right Fig.~\ref{fig:shells}). This is why the Gaussian jet model converges between \textsc{Katu} and \textsc{afterglowpy} at $\sim 10^{3}$ seconds, and a similar effect plays out in the power-law case. If on-axis, the dominant flux from the jet tip will hide this effect.

We should stress that the effect of $M_{\rm ej}$ is an artificial effect imposed by the initial conditions we set on the shell model, and that a different coasting set-up will have a different impact on when each annulus decelerates and thus the flux observed. We discuss this further in section~\ref{sec:M_ej_discuss}.

\section{General IC scattering}
\label{subsec:KNvTM}

\subsection{TeV light curves}
\label{subsec:TeVLC}

In Fig.~\ref{fig:LC_2TeV}, we present the light curves for the 12 scenarios from section~\ref{subsec:Res_12} at 2 TeV. Since the IC regime will mirror the synchrotron regime at a higher energy, the difference between the light curves will be caused by achromatic effects arising from the jet structure and observer angle. The top-hat on-axis light curves will have the highest flux level to start with, due to having a high, consistent energy which will give the most flux when integrating over the beamed solid angle for the observer. Since $E_{0}$ dominates the on-axis flux, the Gaussian and power-law jets are nearly identical to the top-hat case, with only the late-time view of the jet edge revealing the difference in structure. The top-hat jet at $0.5 \theta_{w}$ observer angle is also at the same level as there is no change in the energy across the jet, though due to small numerical effects it appears slightly higher. Any drop-off for this light curve occurs only at the jet edge or beyond, when the nature of the jet is apparent at the start of the run, and the emission is primarily coming from a region where $\mu \neq 1$.

The angle of observation is the most important effect on the light curves. As we go further off axis, we see the TeV light curves exhibit a peak flux whose height decreases and which moves to later times. This is contrasted by the decrease in $M_{\rm ej}$ which is dictated by $E(\theta)$ and how steep the drop-off of $E(\theta)$ is. As discussed in section~\ref{subsec:Res_12}, the lower $M_{\rm ej}$, which arises from the lower energy and fixed $\gamma_{0}$, will lead to an earlier deceleration, which will allow the beaming angle ($1 / \gamma$) around the line of sight to start increasing earlier, marking the turnover of the light curve. Since the Gaussian and power-law jets have an energy drop-off (with the Gaussian profile being steeper), they will see an earlier deceleration off axis and thus an earlier peak flux. Additionally, due to the steeper energy drop-off of the Gaussian, and subsequently less energy over the integrated beaming angle at a given observer angle, its peak flux level will be lower than that of the power-law, which itself will be lower than the peak flux of the top-hat jet with constant energy across angle. In all cases, as the tip of the jet comes into view, we see a convergence in all light curves to the same flux due to $E_{0}$. Once the jet has sufficiently decelerated, $E_{0}$ will no longer dominate the flux and the light curves will begin to diverge again.

\begin{figure*}
    \centering
    \includegraphics[scale=0.35, trim={0.5cm 1.5cm 0 0}]{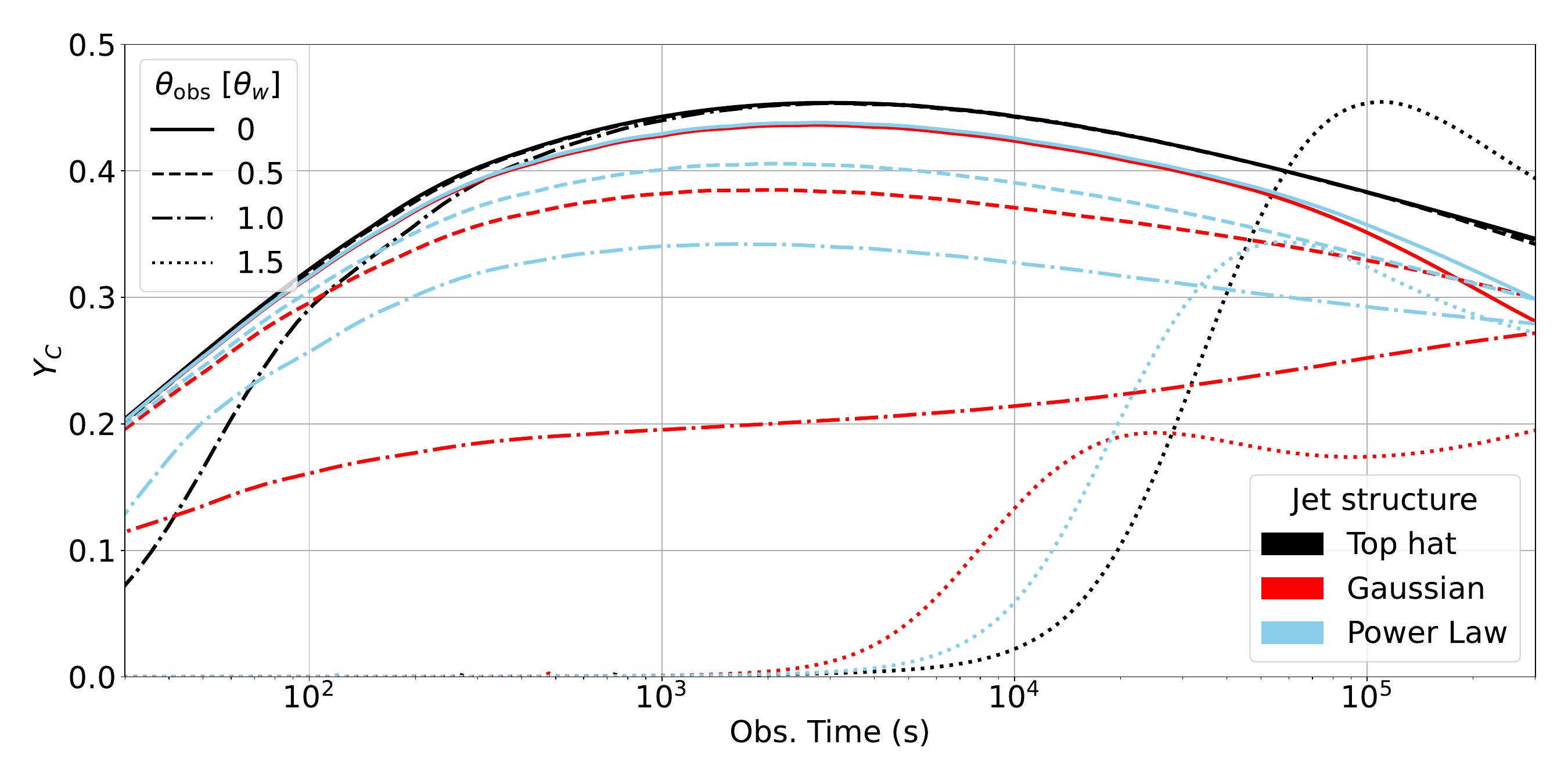}
    \caption{Compton Y-parameter at $\nu_{c}$ ($Y_{C}$) for a series of observer angles (0, 0.5, 1 and 1.5 times the jet width $\theta_{w}$) relative to the jet normal for a top-hat (black), Gaussian (red) and power-law (blue) jet structure. $Y_{C}$ follows a similar relationship to that shown in Fig.~\ref{fig:LC_2TeV}, with the Top-hat jet structure having the highest $Y_{C}$, followed by power-law and Gaussian. Going further off axis has a reducing effect on $Y_{C}$ as well. Parameters are given in table~\ref{tab:12_run_params}).}
    \label{fig:Y_ALL}
\end{figure*}

We stress here again that the off axis light curves heavily depend on the initial coasting phase, which for these runs has an elevated $\gamma_{0}$ to bring out these effects. Furthermore, by fixing $\gamma_{0}$, we have implicitly set the baryon loading of the jet as being angle dependent. This is an arbitrary choice in these runs. Letting $M_{\rm ej}$ be fixed and $\gamma_{0}$ vary with angle leads to a different set of off axis light curves, which we look at in section~\ref{sec:M_ej_discuss}.

While we see that the TeV light curves will be directly impacted by the aforementioned achromatic effects, there is an additional chromatic effect which affects the strength of the IC scattering, and by extension the relative strength of the TeV emission compared to the synchrotron emission. To show this, we make use of the Compton Y-parameter, which is a measure of the strength of the IC scattering with respect to the synchrotron process. This is usually done by finding the ratio of the SSC to synchrotron power in energy space, though at the cooling break $\gamma_{c}$, we can approximate it by taking the ratio of the SSC and synchrotron peaks in $\nu F_{\nu}$,
\begin{equation}
    Y_{C} = \frac{\nu F_{\nu_{c}, SSC}}{\nu F_{\nu_{c}, S}},
\end{equation}
where $Y_{C} \equiv Y(\gamma_{c})$. 

In Fig.~\ref{fig:Y_ALL}, we present the corresponding $Y_{C}$ values over observer time for all observer angles and jet structures. We see that $Y_{C}$ increases and decreases in a similar manner to the TeV light curves in Fig.~\ref{fig:LC_2TeV}; the more energetic top-hat jet structure provides an overall higher peak $Y_{C}$ followed by the power-law and Gaussian jet structures, while going off axis in the power-law and Gaussian scenarios also has a reducing effect on $Y_{C}$ and its peak value. For the top-hat, we see no difference from observer angle, unless at the jet edge or beyond. In this case $Y_{C}$ eventually rises to match the on-axis case as the jet comes into view, most noticeably in the $1.5\theta_{w}$ case where $Y_{C}$ rapidly jumps up to the max $Y_{C}$ value at a later time, before dropping off with the rest of the observer angles.

The power-law and Gaussian jet structures also show a strong correlation with the energetics of the jet at a given annulus and $Y_{C}$, as the further off axis we go, the stronger the impact on reducing $Y_{C}$ throughout the period being observed. An interesting demonstration of this is given by the $1.5\theta_{w}$ Gaussian case, where we see $Y_{C}$ rise as the jet starts to come into view, start to drop off as the jet continues to decelerate, before rising again as the tip of the jet comes into view.

All of this strongly indicates that the IC scattering is dependent on the energetics of the jet, which itself has an angular dependence (if not top-hat). We suggest that this is best explained by considering two competing effects; the \textbf{B}-field, which controls the synchrotron emission, and the rate of electron injection. From the shell model, the \textbf{B}-field is dependent on the internal energy density (see Eq.~\ref{eq:SJ_mid} and Eq.~\ref{eq:B-field}). From this, it follows in the relativistic limit that $B \propto \gamma$. By contrast, the rate of injection is dependent on the Lorentz factor, velocity and radius of the shock (see Eq.~\ref{eq:Inj_rate}), which, in the relativistic limit, gives us $Q_{e,0} \propto \gamma^{8/3}$. This means that the rate of electron injection grows much faster than the \textbf{B}-field, such that for a similar synchrotron photon population we increase the rate of IC upscattering and thus see an increase in $Y_{C}$ and the TeV emission in more energetic parts of the jet.

We should note that the effect in this case (with $E_{0}$ and all other parameters fixed) is small, with $Y_{C}$ only increasing by just over twice as much from the least energetic (Gaussian off axis) to most energetic (top-hat on-axis) case.

\subsection{Impact of different IC cooling regimes}
Both the synchrotron and IC domains in the spectrum will be impacted by how the IC scattering is treated (eg. \citealt{2009ApJ...703..675N, 10.1093/mnras/stab911}). The inclusion of an energy-dependent cross-section, which causes the KN effect, can mitigate IC cooling, but will not entirely remove its impact on the synchrotron spectrum. We can show this in \textsc{Katu} by contrasting three different scenarios: a synchrotron-only run, a run with Thomson cooling (as given in Eq.~\ref{eq:Thom_Cool}) and a run with general IC cooling (henceforth called the KN case). A similar analysis has been performed by \citet{McCarthy_2024}, who take a semi-analytical approach by simplifying the KN cross-section and calculating the Compton Y-parameter at each point in energy space, which is then used along with the spectral breaks to obtain the synchrotron spectrum.

\begin{figure}
    \centering
    \includegraphics[width=\linewidth]{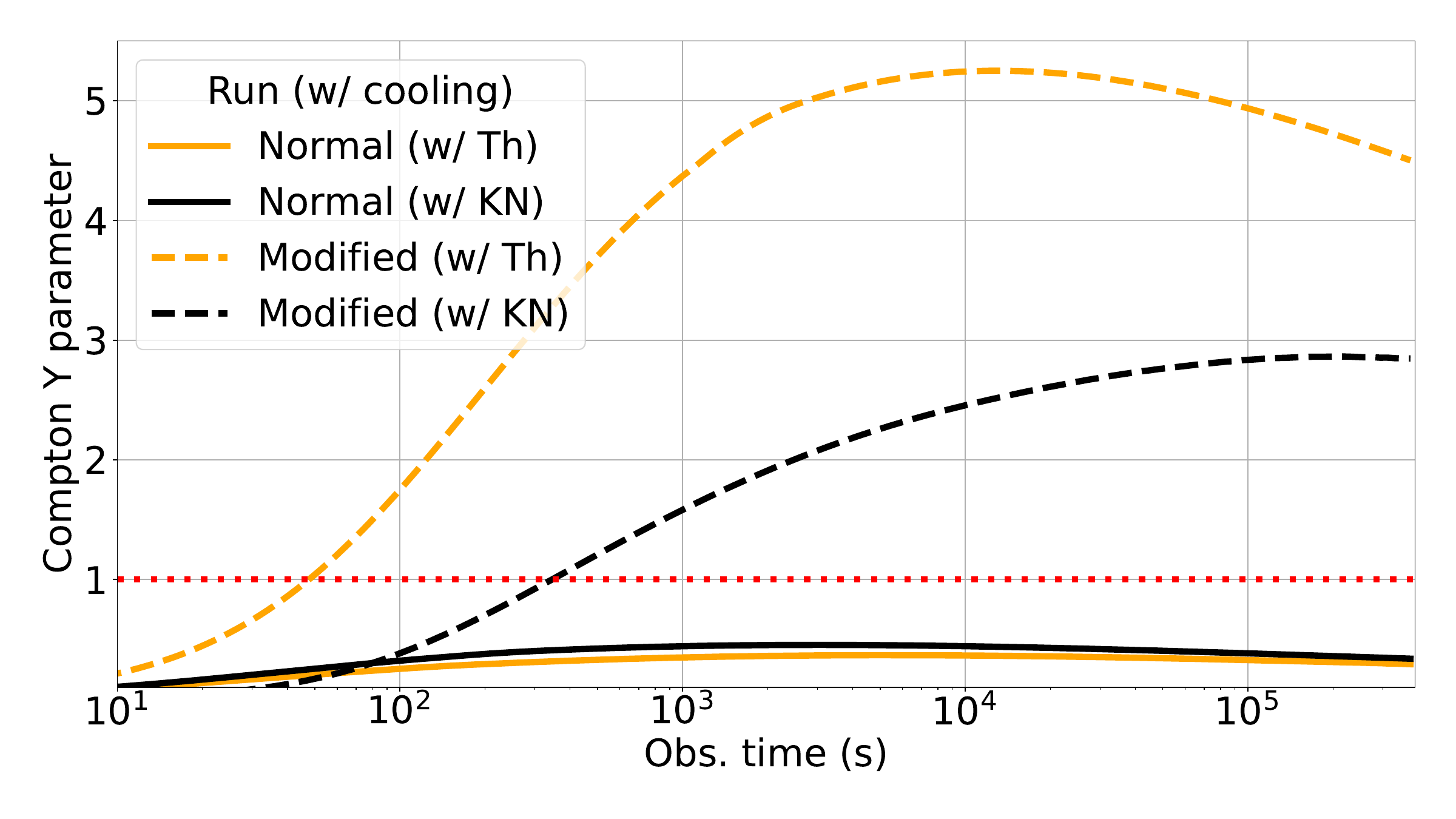}
    \caption{Compton Y parameter over observer time for runs with Thomson and Klein-Nishina corrected cooling for the electrons (photons have the KN-correction enabled in all cases), using the normal parameters (those found in table~\ref{tab:12_run_params}), and the modified parameters (where $\epsilon_{e} = 0.5$ and $\epsilon_{B} = 10^{-4}$). $Y = 1$ is included for reference.}
    \label{fig:Y_param}
\end{figure}

We show the evolution of $Y_{C}$ in observer time in Fig.~\ref{fig:Y_param} for the top-hat, on-axis run from section~\ref{subsec:Res_12} (which here we call the normal run, with $\epsilon_{e} = 10^{-1}$ and $\epsilon_{B} = 10^{-2}$). Because of the low Compton potential  $\kappa_{\rm SC} = \epsilon_{e} / \epsilon_{B}$ (\citealt{derishev2021numerical}), IC scattering is unable to dominate the cooling process, and $Y_{C}$ remains below unity. To allow for the effects of IC scattering to become noticeable, we can increase $\kappa_{\rm SC}$ by raising $\epsilon_{e}$ to $0.5$ and reducing $\epsilon_{B}$ to $10^{-4}$ (the modified run). By doing this, IC scattering can have a more substantial impact around the cooling break, and $Y_{C}$ increases above unity to reflect this.

Since the photon upscattering in \textsc{Katu} already accounts for the KN cross-section, the IC peak obtained in the Thomson cooling case will be lower rather than higher than the KN cooling case, as the KN cooling is weaker at higher energies and so more electrons are present to upscatter photons. This means that in the normal run, the KN case will have a higher Compton Y-parameter, even though in practise we expect the opposite \citep{2009ApJ...703..675N}. However, if we increase the Compton potential, we will increase the number of synchrotron seed photons, as well as the Thomson cooling due to its dependence on the photon energy density (Eq.~\ref{eq:Thom_Cool}). The KN cooling will be affected to a lesser degree as it is also affected by the KN cross section, which drops off at high energies. The result of this is that in the Thomson case, $\nu_{c}$ will be shifted significantly more redward than the KN case, which will also mean there will be more photons at the cooling break to be upscattered to the corresponding cooling break for the IC scattering. This overrides the original effect and causes the Thomson $Y$-parameter to be higher than the KN case in the modified run as in Fig.~\ref{fig:Y_param}. The upshot of this is that Fig.~\ref{fig:Y_param} is able to demonstrate that IC scattering and cooling have become important, but no further conclusions can be obtained from this result. This issue is avoided in the following analysis as we only compare the Thomson and IC cooling in the synchrotron regime, where the cooling term used is more important.

\begin{figure*}
    \centering
    \includegraphics[scale=0.4, trim={2cm 1cm 0cm 0cm}]{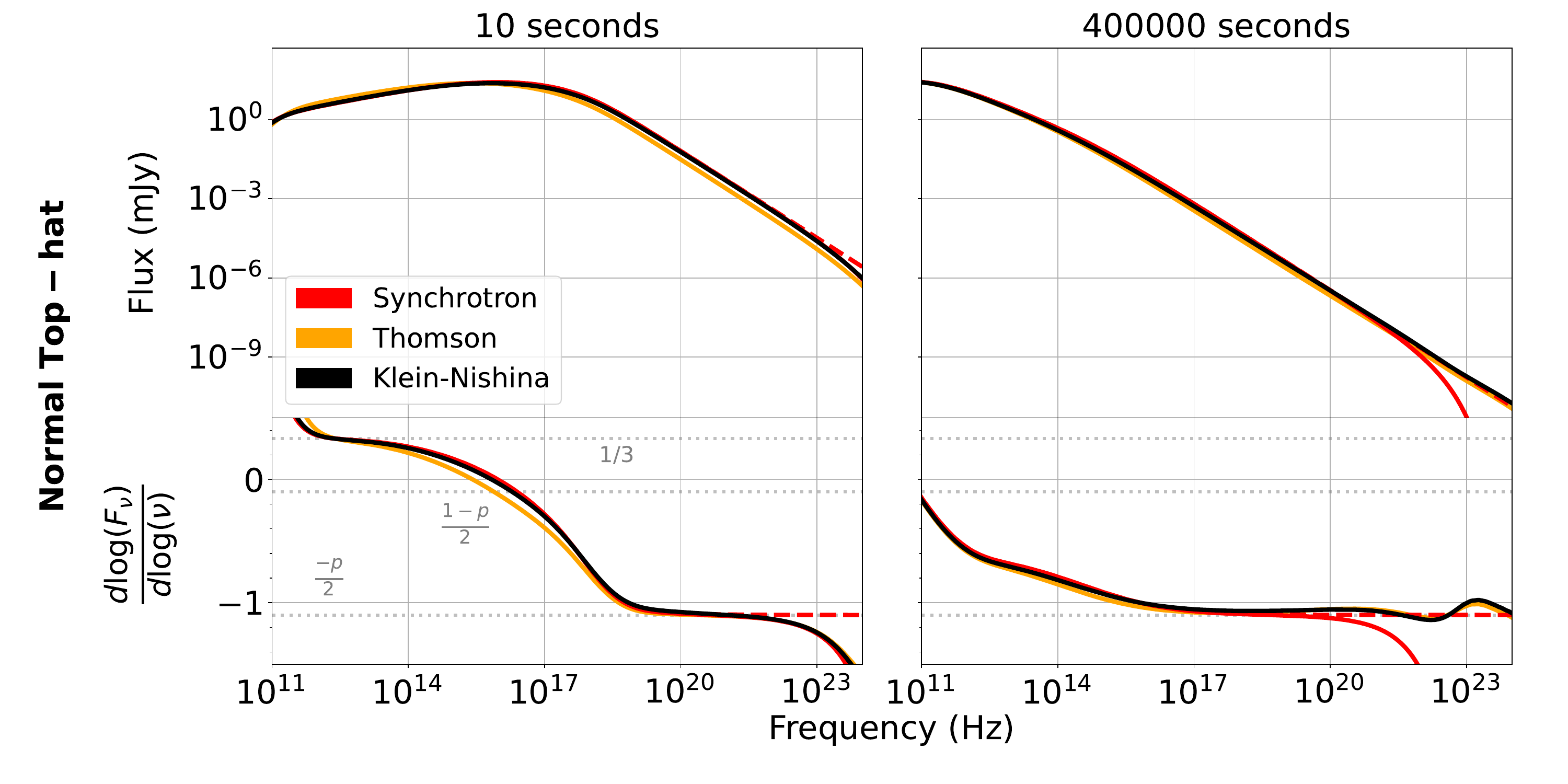}
    \includegraphics[scale=0.4, trim={2cm 1cm 0cm 0cm}]{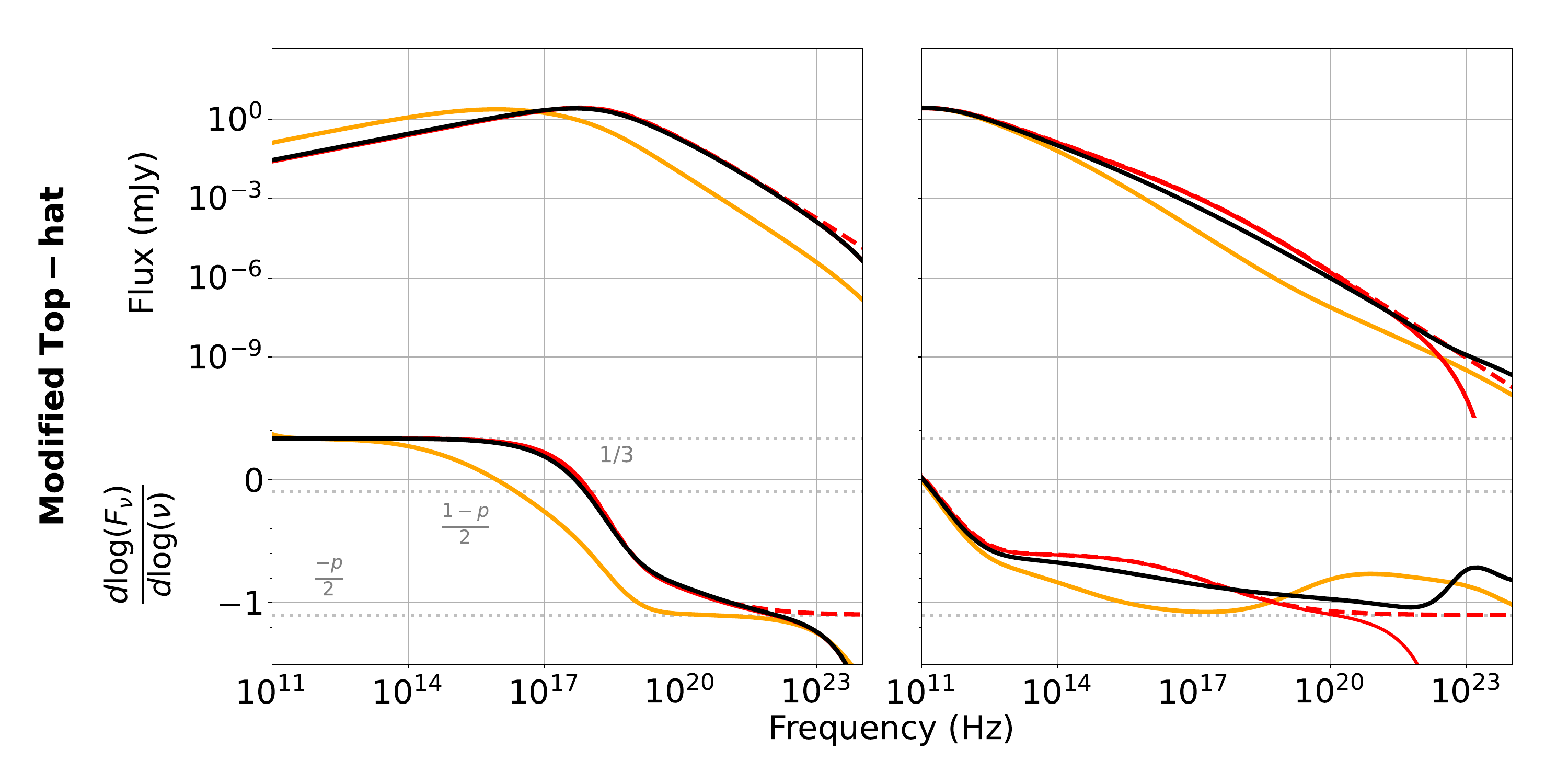}
    \caption{Spectra and spectral gradients for a series of runs with synchrotron-only (red), Thomson IC cooling (blue) and general/KN IC cooling (black), at the start and end of the simulation ($10^{-4}$ days to $4.6$ days). A synchrotron run without a cut-off is also shown (dashed red line). The top row uses the parameters from table~\ref{tab:12_run_params} (the normal run), while the bottom row sets $\epsilon_{e} = 0.5$ and $\epsilon_{B} = 10^{-4}$ to increase the Compton potential (the modified run).}
    \label{fig:TM_plots}
\end{figure*}

\begin{figure*}
\centering
\begin{subfigure}{0.48\linewidth}
    \centering
    \includegraphics[scale=0.22, trim={5cm 2cm 0cm 0cm}]{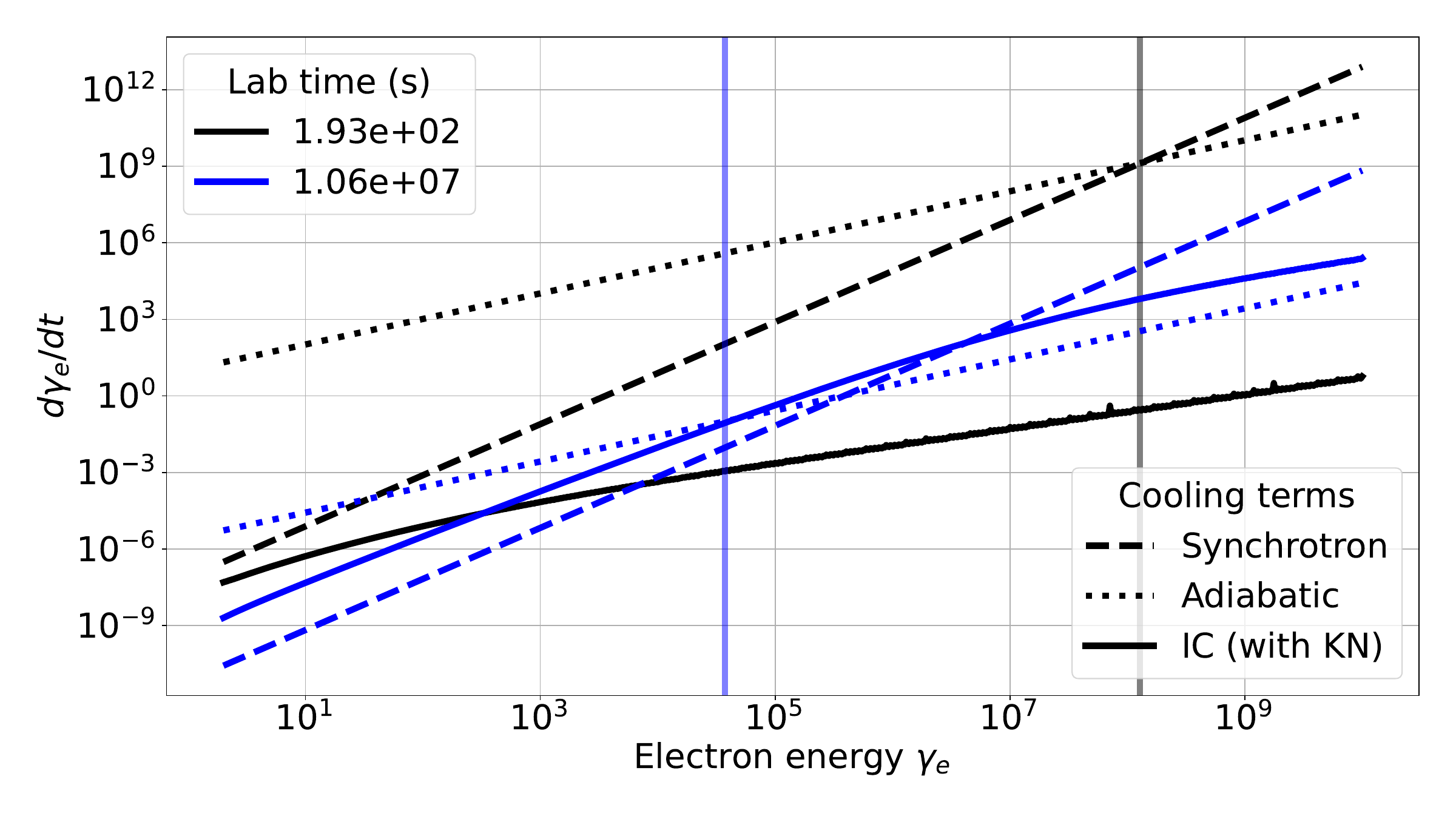}
    \caption{} \label{fig:dgdt}
\end{subfigure}%
\hspace{0.4cm}
\begin{subfigure}{0.48\linewidth}
    \centering
    \includegraphics[scale=0.22, trim={5cm 2cm 0cm 0cm}]{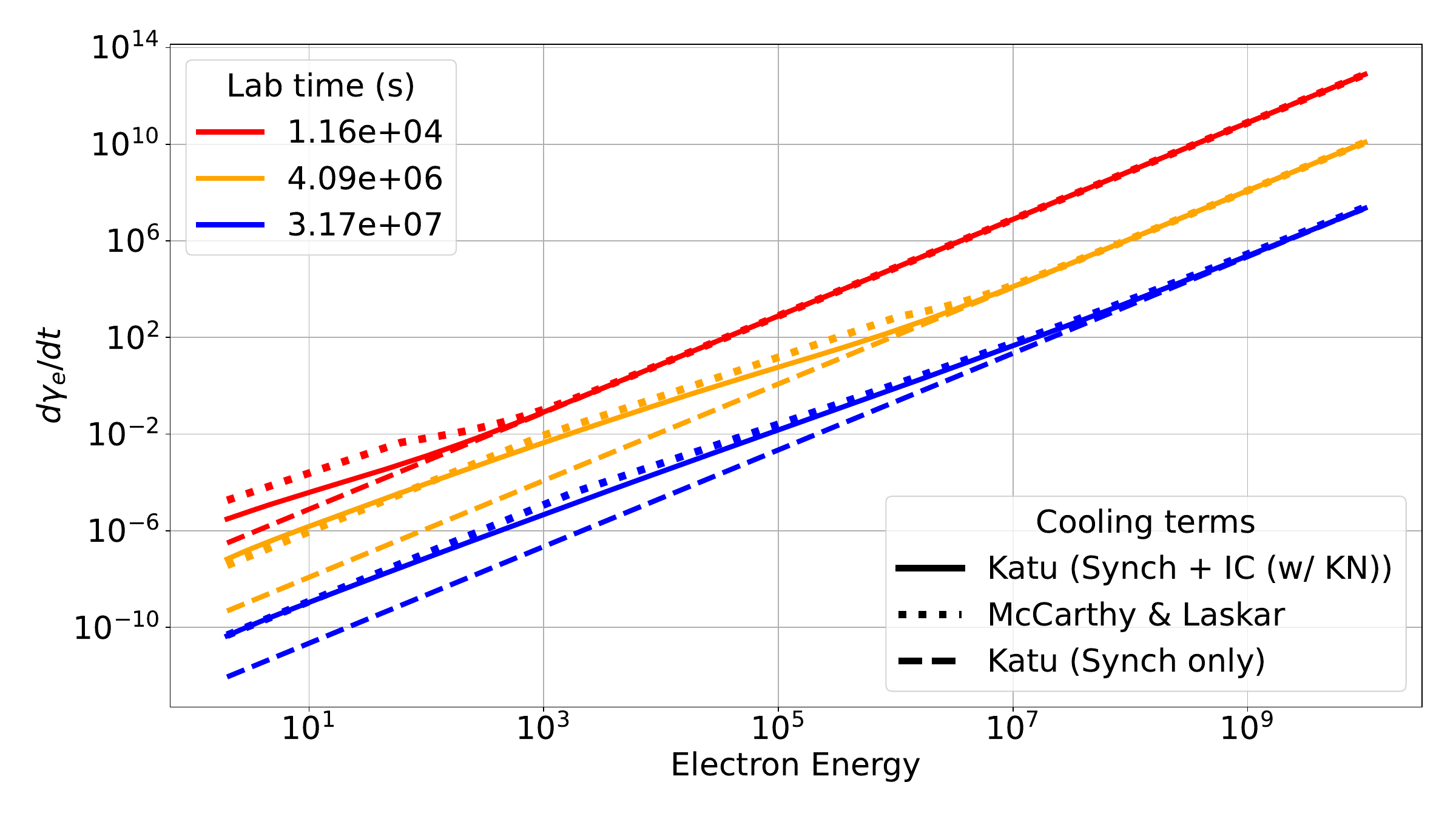} 
    \caption{} \label{fig:dgdt_Comp}
\end{subfigure}%
\caption{Electron cooling in energy space for a series of lab times in the KN corrected modified run. (a) Comparison of the synchrotron [dashed], adiabatic [dotted] and IC cooling (KN corrected) [solid], for an early lab time ($193$ s, black) and a late lab time ($\sim 10^{7}$ s, blue). The position of $\gamma_{c}$ is marked as a translucent vertical line, and is calculated from the turnover of the adiabatic cooling to the synchrotron and IC cooling. (b) Comparison of the combined synchrotron \& IC cooling for \textsc{Katu} (solid) and the Y-parameter model provided by \citet{McCarthy_2024} (dotted). The synchrotron cooling from \textsc{Katu} (dashed) is also included for reference.}
\end{figure*}

The spectra, and corresponding spectral slopes, from these runs are shown in Fig.~\ref{fig:TM_plots}. We show the results from an early time ($\sim 10^{-4}$ days) and a later time ($4.6$ days), with the top rows being the normal runs, and the modified runs on the bottom row. Along with the synchrotron, Thomson, and KN cases, we also show the synchrotron case without the cut-off for reference (the dashed red line). Without a cut-off, the synchrotron spectrum post-cooling break will continue to decrease by $-p / 2$ until it reaches the rightmost energy bin of the simulation. The inclusion of $\gamma_{\rm max}$  will cause a drop-off in the spectrum that will decrease in frequency over time. Since the decrease in $\gamma_{\rm max}$ is related to the \textbf{B}-field (from Eq.~\ref{eq:gamma_max}) as $\gamma_{\rm max} \propto B^{-1/2}$, the change in frequency in the fluid frame $\nu^{'}_{\rm max}$ will be negligible, as $\nu^{'}_{\rm max} \propto \gamma_{\rm max}^{2}B$. Thus, from the observer's perspective, they will see $\nu_{\rm max}$ decrease due to the deceleration of the jet, such that $\nu_{\rm max} \propto \gamma\nu^{'}_{\rm max} \propto t_{\rm obs}^{-3/8}$.

As expected, the normal case shows little difference between the three scenarios, with the only noticeable feature being the cut-off which initially starts at $\sim 10^{22}$ Hz, and by $4.6$ days is at $\sim 10^{20} Hz$. The spectral slopes at early times\footnote{As established near the start of section~\ref{subsec:Res_12}, the normal run starts in fast cooling} show the $1/3$ rising slope up to $\nu_{c}$, before it smoothly transitions across $\nu_{m}$ to $-p/2$. By $4.6$ days, $\nu_{m}$ has moved leftwards such that we see the slope fall to $(1-p)/2$ before gradually transitioning to the $-p/2$ slope. We see in both the Thomson and KN case a slight flattening of the slope from the photons boosted by IC scattering, which then extends the cut-off up to the point where the spectrum transitions from the synchrotron to IC regime. Subsequently, the slope shows a small shallowing as it enters the IC regime.

From the beginning of the modified run, there is a noticeable shift in $\nu_{c}$ of around 2 orders of magnitude between the Thomson run, and the synchrotron and KN runs. This is caused by the highly efficient Thomson cooling compared with KN cooling at the high energies present in the electron population. As this run continues, we start to distinguish between all three scenarios. While they converge at $\nu_{m}$, only the synchrotron slope remains at $(1-p) / 2$ from $\sim 10^{13}-10^{15}$ Hz, before smoothly dropping to $-p / 2$ by $10^{20}$ Hz. The Thomson run shows a more rapid transition across the lower value of $\nu_{c}$ to $-p/2$ by $10^{17}$ Hz, which then flattens into an extended IC peak. By contrast, the KN spectrum shows a gradual transition from $(1-p)/2$ to $-p/2$, due to the minimally affected $\nu_{c}$. The spectrum is extended beyond the cut-off due to IC scattering with inefficiently cooled electrons, before showing a sharper and shallower IC peak beyond $10^{22}$ Hz.

The evolution of the KN case is heavily influenced by its impact on the electron cooling compared to the adiabatic and synchrotron cooling terms. To show this, we present the electron cooling terms for the modified KN run across energy space in Fig.~\ref{fig:dgdt}. For clarity, we show two lab times, which represent an early ($193 $ s, black) and late ($\sim 10^{7}$ s, blue) time in the run. The cooling break is also shown as a vertical line on the plot.

We initially see that the cooling is almost completely dominated by the adiabatic cooling of the population, with synchrotron cooling only overtaking adiabatic cooling at $\gamma_{e} = 10^{8}$. The IC cooling is so weak that it has no impact on the electron cooling, and does not contribute to the positioning of the cooling break. However, as the jet decelerates, the adiabatic and synchrotron cooling drop and the IC cooling starts to rise from the increasing amount of IC scattering at lower energies. Consequently, while the adiabatic cooling is still dominant at low energies, the IC cooling takes over just before $\gamma_{e} = 10^{5}$, as well as setting the position of $\gamma_{c}$. The synchrotron cooling then closely follows the KN cooling before the decreasing the cross-section causes the IC cooling to turnover and the synchrotron cooling begins to dominate around $\gamma_{e} =  10^{7}$. This is why the KN IC cooling only has a noticeable, if small, impact at late times.

We can also compare our cooling to the KN cooling model for GRB afterglows provided by \citet{McCarthy_2024}. Using the prescription and code provided by them in their paper, we can calculate the total synchrotron and IC cooling in their model by finding the KN corrected Y-parameter across energy space, which then modifies the synchrotron cooling:
\begin{equation}
    \frac{d\gamma_{e}}{dt} = -\frac{\sigma_{T}B^{2}}{6\pi m_{e} c}\gamma_{e}^{2}\left(1 + Y(\gamma_{e})\right).
\end{equation}
We have plotted the combined synchrotron and IC cooling from \textsc{Katu} (solid) and from their model (dotted) in Fig.~\ref{fig:dgdt_Comp}. Synchrotron-only cooling is also included from \textsc{Katu} (dashed), though both models have identical synchrotron cooling. Initially, the models diverge from the beginning by around an order of magnitude, with \textsc{Katu} converging more rapidly to the synchrotron dominant regime. As time progresses, we see both models generally agree, with \textsc{Katu} having a smoother transition from the IC dominated to the synchrotron dominated regime, the latest lab time having the longest transition. In contrast, the model provided by McCarthy \& Laskar shows asymptotic relations which show sharp breaks, but which generally agrees with \textsc{Katu}. Overall, we get good agreement between the two models, except at early times low energy, and with some small differences of cooling rates at the transitional phase between the IC and synchrotron cooling regimes.

\begin{figure*}
    \centering
    \includegraphics[scale=0.34, trim={3cm 1cm 4cm 0}]{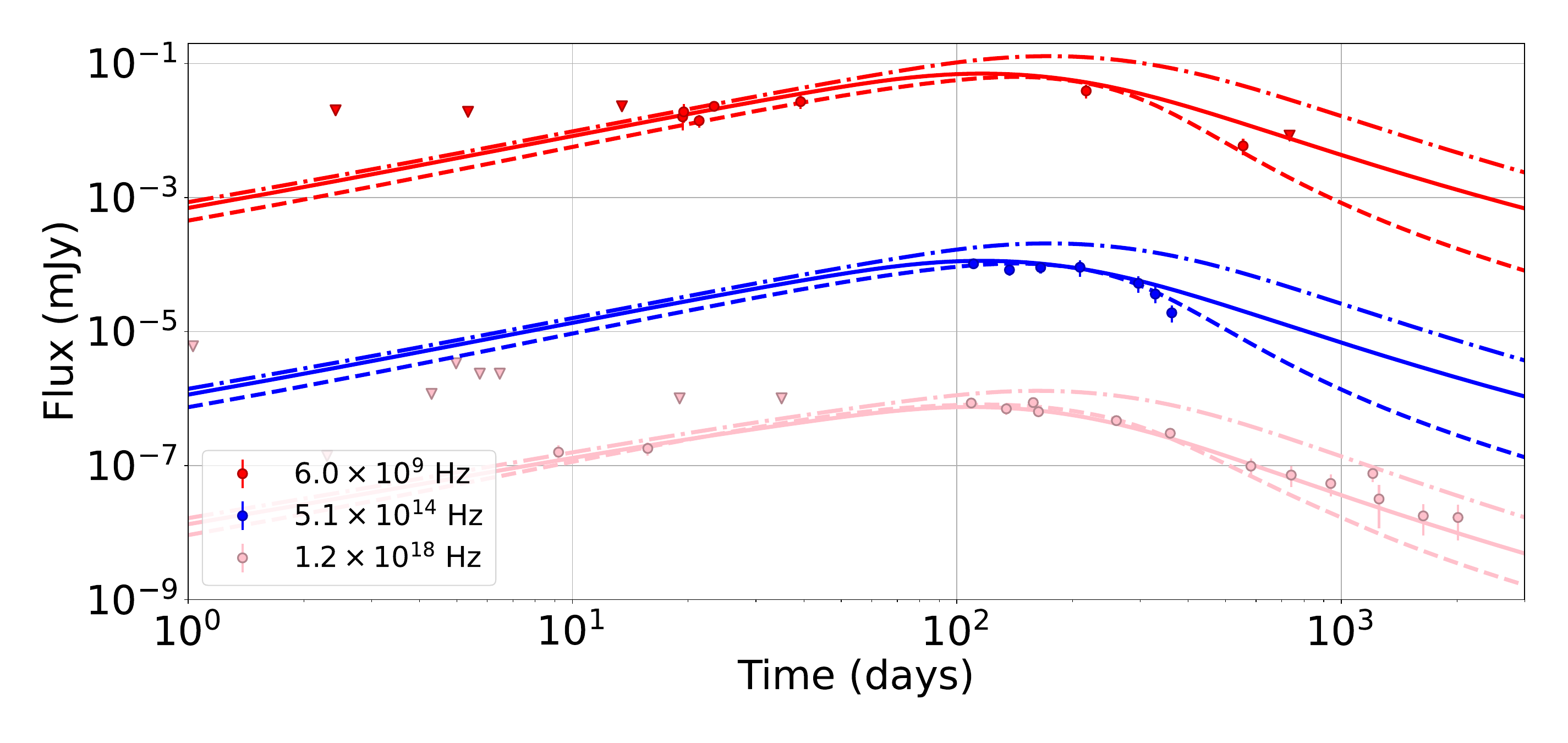}
    \caption{Data points from $1 - 3000$ days \citep{Eyles-Ferris_2024} for GRB170817A for 6 GHz (red), r-band (blue) and X-rays (pink). Overlaid are the best-fit light curves from \textsc{afterglowpy} with spreading from \citet{afterglowpy2} (dashed line), the equivalent result from \textsc{Katu} (dot-dash), and the result from \textsc{Katu} with re-scaling applied (solid). While we appear to get a good late time fit for the X-rays, this is an artificial effect due to the lack of jet spreading in \textsc{Katu}. Observational data obtained from https://gwapa.web.roma2.infn.it/} \label{fig:170817A}
\end{figure*}

\section{Application to GRB 170817A}
\label{subsec:TestCase}
We also want to test our code's ability to reproduce real GRBs. We use GRB 170817A as it is an off axis structured jet with an extensive set of observational data over $3000$ days.

A Bayesian fit using our code would be the ideal way to fit our model to this dataset. However, the high computational cost of calculating such off axis jets makes this currently impractical with \textsc{Katu}, which means that we must turn to an alternative approach. Since \textsc{afterglowpy} uses an almost identical shell model and has been shown to fit well with \textsc{Katu} already, we made use of the best fit of 170817A provided by \citet{afterglowpy2}, where \textsc{afterglowpy} was fitted to 170817A using the \textsc{emcee} package \citep{Foreman-Mackey_2013}. We selected the Gaussian jet, as it was found to have a better overall fit, and ran \textsc{Katu} based off the the parameter set obtained by \textsc{afterglowpy} \footnote{Electron-positron pair production was enabled for this run to allow for a direct comparison with \citet{Clement}, but we found that it did not impact the outcome.}. The results from this run can be found in Fig.~\ref{fig:170817A}, with \textsc{afterglowpy} presented with the dashed line, and \textsc{Katu} with the dot-dashed line.

\begin{table}
    \centering
    \caption{Table of parameters used for the 170817A runs. The normal parameters refer to those provided by \citet{afterglowpy2}, while the re-scaled paramters are those used by \textsc{Katu} for the re-scaled run. Units can be found in table~\ref{tab:12_run_params}.}
    \begin{tabular}{lcc}
    \hline
    Parameter & Normal & Re-scaled \\
    \hline
    $E_{0}$ & $4.8 \times 10^{53}$ ergs & $2.16 \times 10^{53}$ ergs \\
    $\theta_{w}$ & 15.47\textdegree & $\sim$ \\
    $\theta_{c}$ & 3.21\textdegree & $\sim$ \\
    $\theta_{\rm obs}$ & 19.48\textdegree & $\sim$ \\
    $n_{\rm ext}$ & $2.4 \times 10^{-3}$ cm$^{-3}$ & $3.6 \times 10^{-3}$ cm$^{-3}$ \\
    $\gamma_{0}$ & $2000$ & $\sim$ \\
    $R_{\rm fire}$ & $10^{8}$ cm & $\sim$ \\
    z & $0.00973$ & $\sim$ \\
    p & $2.13$ & $\sim$ \\
    $\epsilon_{e}$ & $1.9\times 10^{-3}$ & $1.71 \times 10^{-3}$ \\
    $\epsilon_{B}$ & $5.75 \times 10^{-4}$ & $\sim$ \\
    $\chi_{N}$ & $1$ & $\sim$ \\
    $\eta$ & $1$ & $\sim$ \\
    \hline
    \end{tabular} \label{tab:170817A}
\end{table}

From this first run, we get a light curve which for the first $100$ days shows the same slope as \textsc{afterglowpy}, but obtains an excess of flux in all three bands, once more due to variations in how \textsc{Katu} and \textsc{afterglowpy} determine the location of spectral breaks (as discussed in section~\ref{subsec:Res_12}). Beyond $100$ days, this excess gets worse, missing several observed measurements and exceeding the upper limit of a radio data point at $700-800$ days. The reason for this is that \textsc{afterglowpy} also includes jet spreading, and was fitted with this in mind, while \textsc{Katu}, which is primarily focusing on early-time results, does not. Subsequently, the parameters provided by \citet{afterglowpy2} produce more flux in \textsc{Katu} than otherwise expected in this timeframe.

To correct for both of these effects, and allow \textsc{Katu} to better fit the data itself, the results were re-scaled, taking advantage of the scale in-variance of the parameters (\citealt{vanEerten2012, ScalingRyanEerten}). Once a good fit was found by re-scaling, \textsc{Katu} was re-run with the new parameter set. Both the normal and re-scaled parameters are shown in table~\ref{tab:170817A}, with changes made to $E_{0}$, $n_{\rm ext}$ and $\epsilon_{e}$. The new result (solid line), shows a much better fit to the observed data, though still with some excess beyond $300$ days in the $6$ GHz and r-band, due to the lack of jet spreading in \textsc{Katu}. 

We obtain a much better fit throughout in the X-rays, barring the anomalous data point just after $1000$ days (see e.g. \citealt{Hajela2022, Troja2022X-raypoint} for discussion of this feature). However, beyond $300-400$ days, this is an artificial result arising from the lack of spreading in \textsc{Katu}. Including jet spreading would be physically more accurate, as done by \textsc{afterglowpy}. This also gives a better fit to the 6 GHz and r-band results in the \textsc{afterglowpy} case, but under-estimates the X-ray flux. This is a known problem, and it has been suggested that the cause for the boost in the X-rays is due to a late time injection of energy into the shell, or some other additional component \citep{afterglowpy2}.  

\begin{figure}
    \centering
    \includegraphics[width=\linewidth]{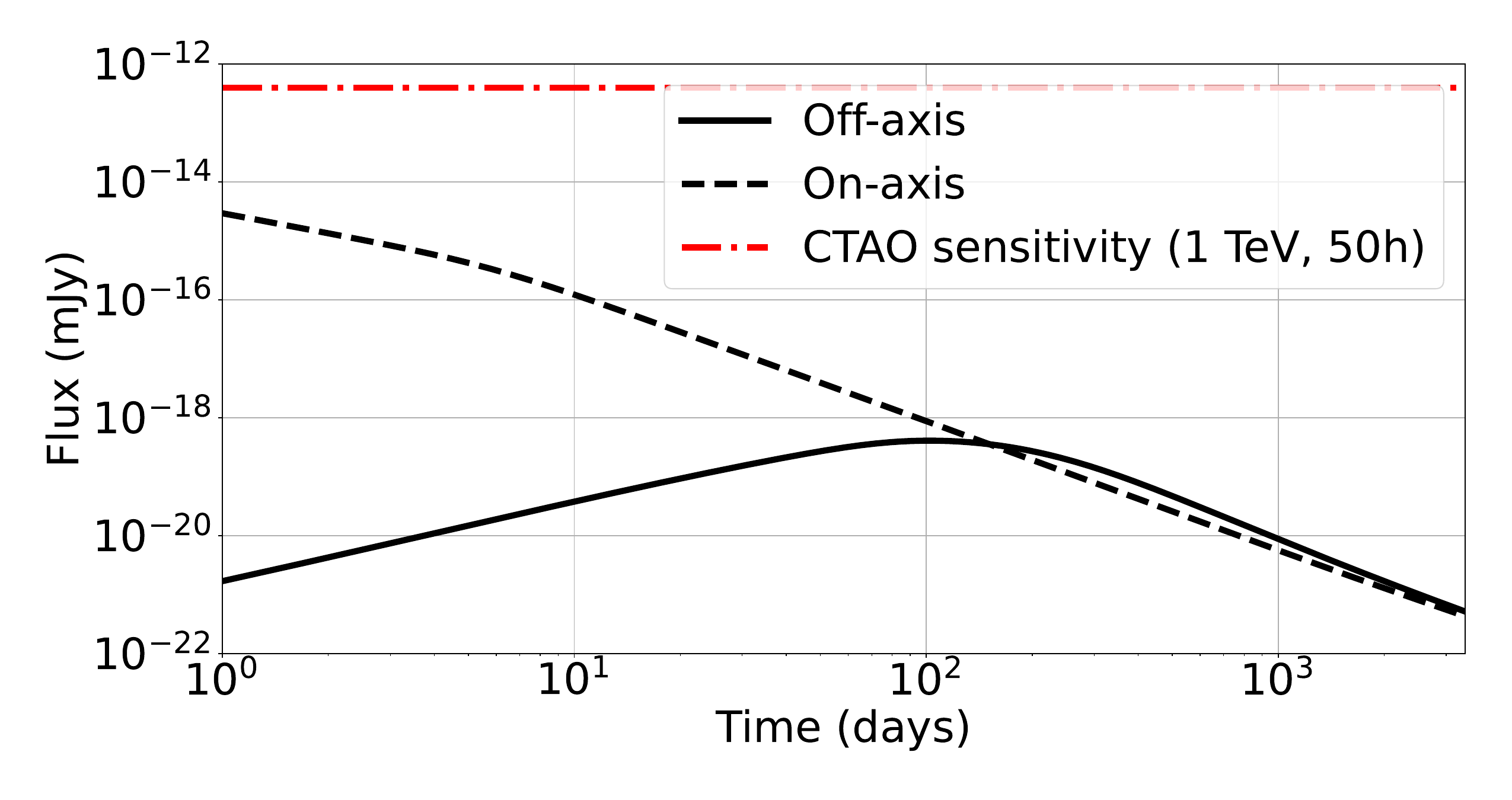}
    \caption{1 TeV light curves for the 170817A re-scaled run, with the off axis observer angle (solid) and on-axis (dashed) shown. The 50h performance for the CTAO at 1 TeV is also provided for reference (red dot-dashed)}
    \label{fig:170817A_TeV}
\end{figure}

\begin{figure*}
    \includegraphics[scale=0.45, trim={0.5cm 1.5cm 0 0}]{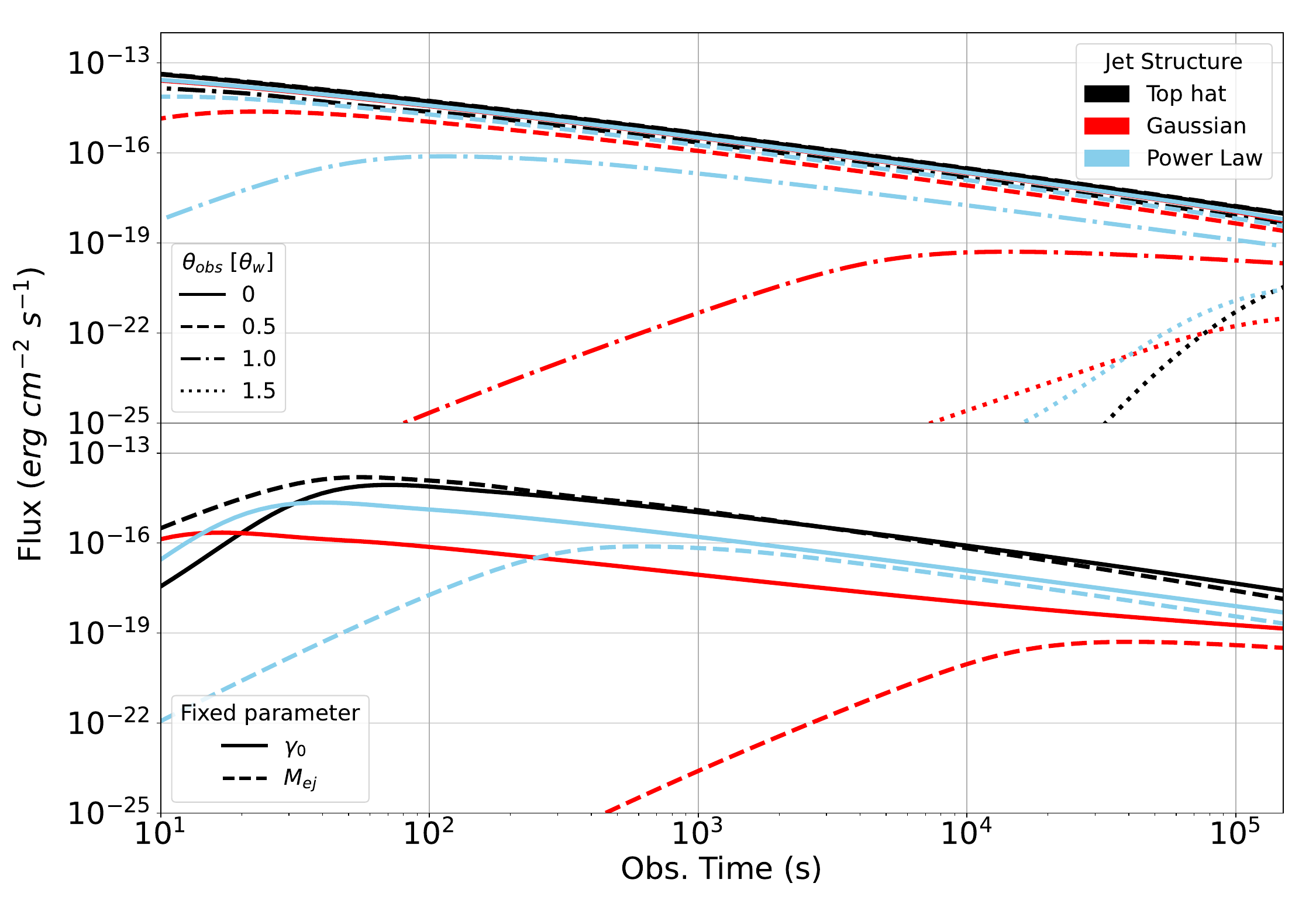}
    \caption{2 TeV light curves where we compare the fixed $\gamma_{0}$ and $M_{\rm ej}$ runs. Jet structure in both plots are given by colour, with top-hat (black), Gaussian (red) and power-law (sky blue). In the top plot, we reproduce a similar figure to Fig.~\ref{fig:LC_2TeV}, but with fixed $M_{\rm ej}$ across angle, instead of fixed $\gamma_{0}$. Observer angle is represented by line style. In the bottom plot, we compare the two types of run for the 3 jet structures, with fixed $\gamma_{0}$ as the solid line, and fixed $M_{\rm ej}$ as the dashed line. Parameters used can be found in table~\ref{tab:12_run_params}, with $M_{\rm ej}$ set to $10^{-4} M_{\odot}$}
    \label{fig:LC_2TeV_2}
\end{figure*}

In Fig.~\ref{fig:170817A_TeV}, we show the 1 TeV light curves for the re-scaled case (solid line), along with the same run but observed on-axis (dashed line). We see the same behaviour of peak turnover due to the difference in observer angle as in Fig.~\ref{fig:LC_2TeV}, which converge at late times due to the same jet and energy structure being present in both cases. Based on the re-scaled fit, we predict that the light curve peaks, as with the other frequency bands, at around $100$ days, with a flux of $4 \times 10^{-19}$ mJy, while the on-axis peaks before $1$ day at a flux level greater than $1.5 \times 10^{-15}$ mJy.

We compare both to the expected 50~hour sensitivity of the upcoming CTAO at 1 TeV\footnote{https://www.ctao.org/for-scientists/performance/}, which acts as the best case scenario. Even with the improved performance over H.E.S.S (\citealt{2019scta.book.....C}), any TeV signature from a comparable GRB like 170817A would not be observed by CTAO, even if on-axis. This would require either an increase in $\kappa_{\rm SC}$ or $E_{0}$ to increase the number of IC scattered photons, though both would have an impact on the fitting of the synchrotron light curves in turn. A possible workaround is to refit with a completely different jet structure and parameter set, which has been recently done by \citet{Clement}, using a semi-analytical model based off \citet{2009ApJ...703..675N} and \citet{10.1093/mnras/stab911}. We discuss this further in the next section.

\section{Discussion}
\label{sec:Discuss}
\subsection{\textsc{afterglowpy} and the effect of $M_{\rm ej}$}
\label{sec:M_ej_discuss}

Overall, we see that \textsc{Katu} works well in simulating the structured jets of GRB afterglows, both in terms of jet structure and off axis measurements. We have used \textsc{afterglowpy} as a reference for testing these features of our code, keeping in mind that there will be variations that arise due to the addition of IC cooling, the shell model including a finite ejecta mass $M_{\rm ej}$, and the use of a radially integrated emission in \textsc{Katu}.

We noted in section~\ref{subsec:Res_12} that the inclusion of $M_{\rm ej}$ caused a variation in the evolution of the early time off axis light curves, due to the fluid starting at a lower but constant $\gamma$ in the shell model. In setting up this model, the assumption was made that initially the bulk Lorentz factor for all annuli would be constant, which as a result of Eq.~\ref{eq:E_ej}, meant that $M_{\rm ej} \propto E_{\rm ej,~iso} \propto E(\theta)$. However, in other models the assumption is normally made that the $\gamma_{0} \propto E(\theta)$ (eg. \citealt{10.1093/mnras/staa538, DuqueJet, 2019MNRAS.489.1820L}), which, if coupled with the initial ejecta energy, means we instead fix $M_{\rm ej} \not\propto \theta$ in the shell model. This can have a substantial impact on the early time light curve of structured jets, where the TeV signal will be strongest.

In Fig.~\ref{fig:LC_2TeV_2}, we show the result of applying this assumption on the 2 TeV light curves from Fig.~\ref{fig:LC_2TeV}. In the top plot, $M_{\rm ej}$ is set to $10^{-4} M_{\odot}$, and if the jet has structured annuli, the initial Lorentz factor is calculated from the varying $E(\theta)$ using Eq.~\ref{eq:E_ej}. The bottom plot compare the jet edge observer angle light curves for both the fixed $\gamma_{0}$ and $M_{\rm ej}$ cases. We observe a small drop in the flux for all on-axis jet structures and the $0.5\theta_{w}$ top-hat jet, though no other impact in the time frame of interest. This is expected, as the on-axis observer will not see information from the coasting phase, and will primarily see the deceleration phase, at which point information about the coasting phase will be lost.

We also see, due to the increased time required for the energy in $M_{\rm sw}$ to overtake the energy in $M_{\rm ej}$ that the turnover in all the light curves is shifted to later times, especially in the jet edge case, where we see a shift from $10$s to $10^{4}$ seconds. This effect is negligible for the top-hat case except when far off axis. The pre-deceleration slopes are also shallower than the ones in Fig.~\ref{fig:LC_2TeV}.

The largest change is the order of the turnovers, which is reversed from the original light curves, and is most evident when the observer is aligned with the jet edge. This is caused by the order of deceleration of the jet annuli being swapped around; because each annulus now has the same $M_{\rm ej}$, they must transfer the same amount of energy to the swept-up matter to decelerate. But since the energy at the tip is higher, and thus its initial Lorentz factor, it is able to more rapidly sweep up the requisite $M_{\rm sw}$, and decelerate first. Conversely, if we fix $\gamma_{0}$ for all annuli, then they will sweep up $M_{\rm sw}$ equally, and since the edge of the jet will have the least $M_{\rm ej}$, it will gain the required energy to decelerate first.

Once the entire jet has begun to decelerate, it will forget the initial conditions, and both scenarios, fixed $M_{\rm ej}$ and fixed $\gamma_{0}$, will converge on each other. However, the further off axis the observer is, the longer it will take for this convergence to occur, and the more  exposed the resultant flux becomes to the initial conditions of the selected model. 

Ultimately, this effect results from our choice of initial baryon loading with angle. Since the off axis light-curve may therefore include information about the coasting phase, we conclude that any future modelling of off axis TeV emission should be mindful of this feature when deciding the baryon loading of the jet.

\subsection{Modelling IC scattering}
\label{subsec:Disc_IC}
Directly comparing the work done in this paper to analytical and semi-analytical approaches (such as \citet{2009ApJ...703..675N, 10.1093/mnras/stab911, McCarthy_2024}) is not trivial. To solve without recourse to numerical solvers, these approaches must apply simplifying assumptions. For example, in the case of \citet{McCarthy_2024}, they took steps such as fixing the synchrotron critical frequency for a given $\gamma_{e}$ as a delta function, and setting the KN cross-section to zero above the Thomson regime rather than modelling the full evolution of the cross-section in energy space. Conversely, a numerical approach may fully treat the range of synchrotron frequencies emitted by $\gamma_{e}$ and the KN cross-section, the result of which is a smoothly changing electron and photon field, rather than approximate power-law asymptotes at given break points.

Subsequently, when we look at the impact on the synchrotron regime in Fig.~\ref{fig:TM_plots}, our KN results shows a smoother transition from the Thomson to KN regime, as compared to the cooling term used by \citet{McCarthy_2024}. We observe that the \textsc{Katu} cooling term begins transitioning at lower energies compared to \citet{McCarthy_2024}, though shortly thereafter their cooling also turns over, and stays within an order of magnitude of the \textsc{Katu} result. The exception to this is the early time result, where we observe an order of magnitude difference before they converge in the synchrotron regime. Once the transition occurs, the KN spectra will converge with the synchrotron result, as in \citet{McCarthy_2024}, though in practise the IC scattered photons prevent this from being observed in the actual observed spectrum.

When looking at the synchrotron spectra between the three cases of synchrotron-only, Thomson and KN cooling in Fig.~\ref{fig:TM_plots}, it is not possible to differentiate between each cooling regime unless there is a high enough Compton potential. If $\kappa_{\rm SC}$ is large, the synchrotron-only and KN cooling are still similar at early times, but we do see a 
noticeable difference in the spectral slopes at later times. In particular, the synchrotron-only slope shows a brief levelling out at $(1 - p)/2$ before dropping down to $-p/2$ and then reaching the cut-off, while the KN slope shows a very short levelling off, followed by a constant decreasing gradient beyond the cut-off, before it reaches the start of the IC component in the spectrum.

\subsection{Application to GRB 170817A}
Since it is the only confirmed off axis GRB with strong indications of structure, 170817A is a good GRB to initially test our full model against, despite the lack of observed TeV emission. As we have shown, this lack of emission does not mean there was no VHE; due to the off axis and weaker energetics of the jet (compare with GRB 221009A in \citealt{221009AStructJet, Ren_2024}), any such TeV emission would not be visible by detectors, including the upcoming CTA observatory.

A similar analysis has recently been performed by \citet{Clement}, using a semi-analytic model built on a jet structure proposed by \citet{DuqueJet}, and building on the KN model from \citet{2009ApJ...703..675N} and \citet{10.1093/mnras/stab911} by including the synchrotron cut-off and pair production. They performed a Bayesian fit using their model, and obtained a TeV light curve, both for the best fit $\theta_{\rm obs}$, and for a series of other angles including near on-axis. They obtained a flux that was elevated compared to our model, and also found that an on-axis emission from 170817A could have been observed by the CTAO. They also show a steepening around the turnover that is absent in our light curve. However, we both roughly agree on the time of turnover, around $100$ days.

As with \citet{McCarthy_2024}, a direct comparison of our models is not trivial. However, the difference in our TeV light curves will come down to variations between our models. Firstly, we made use of a Gaussian jet with a fixed initial Lorentz factor, while the jet from \citet{DuqueJet} applies a power-law structure on both the energy and initial Lorentz factor of the jet. Their fit also obtained a narrower $\theta_{c}$ compared to our jet. Secondly, the fit values obtained for the microphysics differ substantially from ours, with $\epsilon_{B}$ being around $3$ times higher, while $\epsilon_{e}$ was around 11 times higher. This would lead to a higher $\kappa_{\rm SC}$ and thus boosted IC photon field.

We do however agree that the TeV light curve will follow the observed achromatic time evolution of the synchrotron light curves, but at a much lower flux level. Due to the lack of information from the unobserved IC peak, which may have helped constrain $\epsilon_{B}$, the difference in numerical and analytical assumptions for KN-corrected IC modelling, and the different modelling assumptions for the jet structure, we find that whether the TeV light curve of GRB 170817A would have been detectable on-axis remains inconclusive.

\section{Conclusions}
\label{sec:Conc}

In this paper, we make use of a modified version of the blazar kinetic code \textsc{Katu}, which was modified to include adiabatic expansion and fully KN-corrected IC cooling, on top of synchrotron emission, Thomson cooling and pair production. It was further modified to run multiple concurrent simulations to represent structured jet annuli, with independent shell models to calculate the injection parameters and \textbf{B}-field. We also included a means to calculate the flux for both on and off axis jets from the photons fields of these simulations. EBL attenuation was added to account for GRBs whose VHE emission might be impacted by CMB radiation. From this code, we reach the following conclusions:

\begin{itemize}
    \item \textbf{Comparison with \textsc{afterglowpy}} -- A series of on-axis and off axis structured jets are compared against the synchrotron model provided by \textsc{afterglowpy}. We find a generally good agreement between the results in the synchrotron regime, with some variation due to the differing positions of the spectral breaks caused by several effects, including the impact of radially integrating over the shell to obtain the flux. off axis results from \textsc{katu} show more flux early on in the runs during coasting, due to our inclusion of ejecta mass into the model. Here, we also show that the treatment of the initial ejecta mass distribution with angle can have a large impact on the early time light curves, impacting the amount of flux at the peak turnover and time where this turnover occurs.
    \item \textbf{TeV light curves} -- We also obtained TeV light curves for the same series of runs when comparing to \textsc{afterglowpy}. The TeV light curves showed the same achromatic effects from structure and observer angle, with the peak flux level and time it was observed being sensitive to both choices. However, an additional effect from the structure and observer angle was observed in the IC scattering, due to the fact that IC emission was boosted relative to the synchrotron emission closer to the tip of the jet, which we suggested was due to the increased rate of electron injection.

    We also showed that changing the baryon loading so that it was fixed across angles instead of $\gamma_{0}$ caused the observed time of the peak TeV fluxes to reverse, and occur at later times compared to the fixed $\gamma_{0}$ case.

    \item \textbf{IC cooling regimes} -- By comparing our numerical KN-corrected solution for the cooling of electrons to electron Thomson cooling and synchrotron-only spectra (we treat photons using the full KN cross-section throughout), we find noticeable differences in the synchrotron spectrum if the Compton potential, and by extension the Compton Y-parameter, is high, though between the KN-corrected spectrum and synchrotron-only spectrum the only key difference is in the spectral slope, not the flux level. We also find a smoother, albeit similar, transition between the Thomson and KN regimes than in the analytical model presented by \citet{McCarthy_2024}, as the numerical approach is not reliant on asymptotic solutions of the Compton Y-parameter, but instead calculates the cooling from the smoothly changing KN cross-section.
    \item \textbf{Application to GRB 170817A} -- We have also applied our model to a series of light curves from GRB 170817A, using the best fit parameters from \textsc{afterglowpy} \citep{afterglowpy2}, along with with re-scaling of $\epsilon_{e}$, $\epsilon_{B}$ and $E_{0}$ (as outlined in \citealt{vanEerten2012, ScalingRyanEerten}) to improve the match with \textsc{katu}. We confirm a good match, though obtain an excess of late time flux due to the lack of jet spreading in our model. We also get a corresponding TeV light curve for GRB 170817A, which we compare to that found from the model used by \citet{Clement}. We find good agreement on the turnover of the light curve (as an effect of the jet structure), though obtain a lower peak flux by $\sim2$ orders of magnitude, which would not be visible to the CTAO, even if viewed on-axis. We note that, even while both studies show agreement on the off axis flux levels in the synchrotron regime, our values for $\epsilon_e$ and $\epsilon_B$ differ substantially, which will account for part (but not all) of the difference between our studies in the TeV range.
\end{itemize}

With the advent of better multi-messenger probes, and with the CTA observatory soon to come online, there is a strong impetus to develop kinetic models which include jet structure, off axis observation and self-consistent IC scattering, to allow us to better understand the VHE emission of structured-jet GRBs, and to help us further constrain the underlying structure and micro-physics that drive these events.

\section*{Acknowledgements}
JPH would like to thank Geoff Ryan, for helpful discussions on implementing the flux calculations in this work, as well as providing advice on how to compare the \textsc{Katu} and \textsc{afterglowpy} shell models. JPH also thanks Clement Pellou\'in and Tanmoy Laskar for additional discussion.
HJvE, PS and SK acknowledge support by the Science and Technology Facilities Council (STFC) through grant ST/X001067/1, and JPH through grant ST/W507301/1. HJvE further acknowledges support by the European Union Horizon 2020 programme under the AHEAD2020 project (grant agreement number 871158). This work used the Isambard 2 UK National Tier-2 HPC Service (http://gw4.ac.uk/isambard/) operated by GW4 and the UK Met Office, and funded by EPSRC (EP/T022078/1).

\section*{Data Availability}
No new data was generated or analysed in support of this research.




\bibliographystyle{mnras}
\bibliography{refs}




\appendix

\section{Derivation of scaled injection term and adiabatic expansion}
\label{AppAd}
The GRB afterglow blast wave propagates into an external environment. The shell of shocked plasma both expands and sweeps up an increasing amount of external medium. To add adiabatic expansion and its corresponding cooling effect, we need to modify the Boltzmann kinetic equation given by Eq. \ref{eq:KineticEq}. The effects of expansion of the existing electron population and of injection of electron from newly swept-up plasma induce a change over time of the form
\begin{equation}
\label{eq:Boltz}
    \left. \frac{\partial n_{e}}{\partial t} \right|_{\textrm{adb, inj}} = \skew{-8}\dot{n_{e}} - \skew{-6}\dot{\gamma_{e}}\frac{\partial n_{e}}{\partial \gamma_{e}}.
\end{equation}
Here, we have absorbed an injection term $Q_e$ from newly shocked plasma into $\skew{-8}\dot{n_{e}}$ (for a shock wave in a coasting stage, where $\skew{-8}\dot{n_{e}} = 0$ as dictated by the shock-jump conditions, the separate contributions from adiabatic expansion and injection will cancel exactly). 

We therefore need expressions for the adiabatic expansion and injection contributions to $\skew{-8}\dot{n_{e}}$ and $\skew{-6}\dot{\gamma_{e}}$. To obtain these, consider the total number density and internal energy density of the post-shock electrons:
\begin{align}
    &\int_{1}^{\infty} n_{e}(\gamma_{e})d\gamma_{e} = n, \\
    &\int_{1}^{\infty} n_{e}(\gamma_{e})\gamma_{e}m_{e}c^{2}d\gamma_{e} = \epsilon_e e.
\end{align}
Assuming the fluid to obey a polytropic gas law fixes a relation between $n$ and $e$ of the form
\begin{equation}
    \frac{e}{e_{0}} = \left(\frac{n}{n_{0}}\right)^{\hat{\gamma}},
\end{equation}
relative to a starting point $n_0$, $e_0$ and where $\hat{\gamma}$ is the adiabatic index, which we assume to be fixed for this derivation\footnote{For a relativistic population of accelerated electrons, we have $\hat{\gamma} = 4/3$; given the presence of synchrotron radiation, the shock-accelerated electron population is assumed to remain relativistic even after the shock itself has crossed into the trans-relativistic regime.}. Differentiating each side, and noting that $de = n_{e}m_{e}c^{2}\gamma_{e} \epsilon_e^{-1}d\gamma_{e}$ and $dn = n_{e}d\gamma_{e}$, we can find that:
\begin{equation}
    e = \frac{\gamma_{e}m_{e}c^{2}}{\hat{\gamma}\epsilon_e}n.
\end{equation}

Since this is true at all times, we can then find that
\begin{align}
    & \frac{e}{e_{0}}  = \frac{\gamma_{e}}{\gamma_{e,0}}\frac{n}{n_{0}} \\
    \implies & \frac{\gamma_{e}}{\gamma_{e,0}}  = \left(\frac{n}{n_{0}}\right)^{\hat{\gamma} - 1} \label{eq:frac_1}\\
    \implies & \skew{-6}\dot{\gamma_{e}} = (\hat{\gamma} - 1)\gamma_{e}\frac{\dot{n}}{n},
\end{align}
where in the last step we differentiate with respect to time to get $\skew{-6}\dot{\gamma_{e}}$, the adiabatic cooling term.

Returning to Eq.~\ref{eq:frac_1}, if we differentiate with respect to $n$, we can find that
\begin{equation}
    n_{e} = \frac{n_{0}}{\gamma_{e,0}}\left(\frac{n}{n_{0}}\right)^{2 - \hat{\gamma}}\frac{1}{\hat{\gamma} -1},
\end{equation}
from which, if we once again note this is true at all times, we can find that
\begin{equation}
\label{eq:n_e_relate}
    \frac{n_{e}}{n_{e,~0}} = \left(\frac{n}{n_{0}}\right)^{2 - \hat{\gamma}}.
\end{equation}

Taking the time derivative of Eq.~\ref{eq:n_e_relate} in the same manner as for Eq.~\ref{eq:frac_1} then gives us
\begin{equation}
    \skew{-8}\dot{n_{e}} = (2 - \hat{\gamma})n_{e}\frac{\dot{n}}{n}.
\end{equation}

Applying these terms to Eq.~\ref{eq:Boltz} gives us
\begin{align}
    \frac{\partial n_{e}}{\partial t} &= \frac{\hat{\gamma} - 2}{\tau_{\rm ad}}n_{e} - \frac{1 - \hat{\gamma}}{\tau_{\rm ad}}\gamma_{e}\frac{\partial n_{e}}{\partial \gamma_{e}}.
\end{align}
where $\tau_{\rm ad} = -n / \dot{n}$ is the adiabatic timescale in the fluid frame. This can be reorganized in the form of  Eq.~\ref{eq:AdExp}:
\begin{align}
    \frac{\partial n_{e}}{\partial t} &= -\frac{n_e}{\tau_{\rm ad}} - \frac{1 - \hat{\gamma}}{\tau_{\rm ad}}\frac{\partial }{\partial \gamma_{e}} \left( n_{e} \gamma_e \right).
\end{align}

To obtain $\tau_{\rm ad}$, as well as the injection scaling, we must consider how $n$ changes over time. This includes two effects; the injection of newly added and accelerated electrons, and the adiabatic expansion of the fluid. At any moment, the change in $n$ is
\begin{equation}
    \label{eq:A11}
    \Delta n = \frac{N + \Delta N}{V + \Delta V} - \frac{N}{V},
\end{equation}
where $N$ is the total number of particles and $V$ is the volume of the fluid. Noting that for $V >> \Delta V$,
\begin{equation}
    \frac{1}{V + \Delta V} \sim \frac{1}{V}\left(1 - \frac{\Delta V}{V}\right),
\end{equation}
we can rearrange~\ref{eq:A11} to find
\begin{align}
    \Delta n &= \frac{\Delta N}{V} - \frac{N}{V}\frac{\Delta V}{V} \equiv \Delta n_{\rm inj} + \Delta n_{\rm ad},
\end{align}
where we relate the RHS terms to the injection and adiabatic components respectively.

In this context, the volume is found by considering the shell model: if $\Delta R << R_{\rm sh}$, we can approximate the volume to be
\begin{equation}
    \label{eq:V_s}
    V^{'} = 4\pi R_{\rm sh}^{2}\Delta R,
\end{equation}
where $\Delta R$ is the shell width as stated in Eq.~\ref{eq:Delta_R} (and \emph{not} the change in radius of the shock front, which is denoted by $\Delta R_{\rm sh}$), and the prime is used here to denote the lab frame. 

With this in mind, we can find the injection scaling term\footnote{$Q_{e, 0}$ is a Lorentz-invariant.} $Q_{e, 0}$, which is the time derivative of $n_{inj}$:
\begin{align}
    Q_{e, 0} &= \frac{\Delta n_{\rm inj}}{\Delta t} 
             = \frac{\Delta N}{V^{'}\Delta t} 
             = 4(3 - k)\gamma^{2}n_{\rm ext}\frac{\beta_{\rm sh}c}{R_{\rm sh}},
\end{align}
where we consider $\Delta N$ to be the number of particles swept up by the shell over a distance $\Delta R_{\rm sh}$, in a similar manner to Eq.~\ref{eq:V_s}.

We now turn to the adiabatic term. The change in volume is occurring over a time interval where the fluid is decelerating. To express the problem in an inertial reference frame, we can use the shell volume in the lab frame,
\begin{equation}
    V^{'} = \frac{V}{\gamma},
\end{equation}
The fractional change in volume is then
\begin{align}
    \frac{\Delta V}{V} &= \frac{\Delta \left(\gamma V^{'}\right)}{\gamma V^{'}}
                       = 3\frac{\Delta R_{\rm sh}}{R_{\rm sh}} - \frac{\Delta \gamma}{\gamma},
\end{align}
which means that our adiabatic term is
\begin{equation}
    \Delta n_{\rm ad} = -n\left(3\frac{\Delta R_{\rm sh}}{R_{\rm sh}} - \frac{\Delta \gamma}{\gamma}\right).
\end{equation}
The adiabatic timescale is then given by
\begin{align}
    \frac{1}{\tau_{\rm ad}^{'}} &= \frac{\Delta n_{\rm ad}}{n\Delta t}
              = 3\frac{\skew{-11}\dot{R_{\rm sh}}}{R_{\rm sh}} - \frac{\dot{\gamma}}{\gamma},
\end{align}
which in the fluid frame is
\begin{equation}
    \frac{1}{\tau_{\rm ad}} = 3\gamma\frac{\skew{-11}\dot{R_{\rm sh}}}{R_{\rm sh}} - \dot{\gamma}. 
\end{equation}

\section{Updating the Kinetic solver}
\label{AppKE}
Two changes were made to the kinetic equation. The first was to include the adiabatic expansion from appendix~\ref{AppAd}, and the second was to generalise the IC electron cooling so as to include Klein-Nishina effects. The derivation here is similar to the one in \citet{Katu2}, with the addition of these terms.

The original form of the kinetic equation is given by
\begin{equation}
    \frac{\partial n_{e}}{\partial t} = Q_{e} + Q_{i} + \mathcal{L} - \frac{n_e}{\tau_{\rm esc}} - \frac{n_{e}}{\tau_{\rm dec}} - \frac{\partial}{\partial\gamma_{e}}\left(\frac{d\gamma_{e}}{d t}n_{e}\right),
\end{equation}
where $Q_{e}$ is the external injection of electrons, $Q_{i}$ is the internal injection by processes (eg. pair production), $\mathcal{L}$ is the loss of particles by internal processes, $\tau_{\rm esc}$ is the escape timescale, $\tau_{\rm dec}$ is the decay timescale, and the last term is the continuous gains and losses of energy for the electrons due to processes such as synchrotron radiation and IC scattering.

For the purpose of adding adiabatic expansion to \textsc{Katu}, we added to this equation the adiabatic expansion we found in appendix~\ref{AppAd}, namely Eq.~\ref{eq:AdExp}. Doing this modifies the equation to
\begin{multline}
    \frac{\partial n_{e}}{\partial t} = Q_{e} + Q_{i} + \mathcal{L} -\frac{n_{e}}{\tau_{\rm ad}} - \frac{n_e}{\tau_{\rm esc}} - \frac{n_{e}}{\tau_{\rm dec}} \\ - \frac{\partial}{\partial\gamma_{e}}\left(\frac{d \gamma_{e}}{d t}n_{e} + (1 - \skew{2}\hat{\gamma})\frac{1}{\tau_{\rm ad}}n_{e}\gamma_{e}\right).
\end{multline}
The new implementation of Compton cooling can be placed in $\frac{d \gamma_{e}}{d t}$, which in full is
\begin{equation}
    \frac{d \gamma_{e}}{d t} = \frac{1}{\tau_{\rm acc}}\gamma_{e} + S\gamma_{e}^{2} + \frac{d\gamma_{\rm ic}}{dt},
\end{equation}
where $\tau_{\rm acc}$ is the acceleration timescale, S is the proportionality constant for synchrotron losses, and $\frac{d\gamma_{\rm ic}}{dt}$ is the general IC losses as given in equation~\ref{eq:IC_gen_cool}, which replaces the original Thomson scattering pre-factor added to $S$.

Substituting this in, and merging terms, we can obtain
\begin{equation}
      \frac{\partial n_{e}}{\partial t} = Q + \mathcal{L} + An_{e} + B\frac{\partial n_{e}}{\partial \gamma_{e}},
\end{equation}
where
\begin{align}
  Q &= Q_{e} + Q_{i}, \\
  A &= \frac{\skew{3}\hat{\gamma} - 2}{\tau_{ad}} -\frac{1}{\tau_{\rm esc}} - \frac{1}{\tau_{\rm dec}} - \frac{1}{\tau_{\rm acc}} - 2S\gamma_{e} - \frac{\partial}{\partial \gamma_{e}}\left(\frac{d\gamma_{\rm ic}}{dt}\right) , \\
  B &= -\frac{\gamma_{e}}{\tau_{\rm acc}} - S\gamma_{e}^{2} -\frac{1 - \skew{3}\hat{\gamma}}{\tau_{\rm ad}}\gamma_{e} - \frac{d\gamma_{\rm ic}}{dt}.
\end{align}

We note that $\left(\hat{\gamma} - 2\right)$ reappears for $n_{e}$ as $\gamma_{e}$ goes outside the partial derivative on the rightmost side.

Letting $\mathcal{L} = Ln_{e}$, and converting the equation to log space, we can finally obtain the forms used in the \textsc{Katu} solver for charged particles, namely
\\
\begin{align}
  \frac{\partial n_{e}}{\partial t} &= Q + A^{'}n_{e} + B^{'}\frac{\partial n_{e}}{\partial \ln{\gamma_{e}}}, \\
  \frac{\partial \ln{n_{e}}}{\partial t} &= \frac{Q}{n_{e}} + A^{'} + B^{'}\frac{\partial \ln{n_{e}}}{\partial \ln{\gamma_{e}}},
\end{align}
where
\begin{align}
    A^{'} = L + \frac{\skew{3}\hat{\gamma} - 2}{\tau_{\rm ad}} &-\frac{1}{\tau_{\rm esc}} - \frac{1}{\tau_{\rm dec}} \\ &- \frac{1}{\tau_{\rm acc}} - 2S\gamma_{e} - \frac{1}{\gamma_{e}}\frac{\partial}{\partial \ln\gamma_{e}}\left(\frac{d\gamma_{\rm ic}}{dt}\right),
\end{align}
and
\begin{align}
  B^{'} &= -\frac{1}{\tau_{\rm acc}} - \frac{1 - \skew{3}\hat{\gamma}}{\tau_{\rm ad}} - S\gamma_{e} - \frac{1}{\gamma_{e}}\frac{d\gamma_{e}}{dt}.
\end{align}
Since we are only considering electrons for the simulations in this paper, we do not need to modify the solver for neutral particles. 

\section{Glossary of physical terms}
We outline here the mathematical symbols used in this paper, and the frame of reference they are in, if not otherwise stated.

\begin{table}
    \begin{tabular}{lll}
    \hline
    Symbol & Parameter name & Frame \\
    \hline
    $E_{0}$ & Normalised jet energy & Lab \\
    $E_{\rm iso}$ & Isotropic equivalent energy & Lab \\
    $M_{\rm ej}$/$M_{\rm sw}$ & Ejecta/Swept-up mass & Lab \\
    $\theta_{w}$ & Jet half-opening angle & Lab \\
    $n_{0}$ & External number density & Lab \\
    $\gamma_{0}$ & Initial Lorentz factor of jet & Lab \\
    $R_{\rm fire}$ & Initial fireball radius & Lab \\
    $z$ & Cosmological redshift & Observer \\
    $p$ & Spectral index & Fluid \\
    $\epsilon_{e}$ & Electron energy fraction & Fluid \\
    $\epsilon_{B}$ & \textbf{B}-field energy fraction & Fluid \\
    $\chi_{N}$ & Fraction of electrons accelerated & Fluid \\
    $\eta$ & Ignorance parameter for $\gamma_{\rm max}$ & Fluid \\
    $\theta_{c}$ & Jet core width & Lab \\
    $b$ & Power-law index & Lab \\
    $\gamma$ & Bulk fluid Lorentz factor & Lab \\
    $\Gamma_{\rm sh}$ & Shock Lorentz factor & Lab \\
    $\beta$ & Normalised velocity of fluid (to $c$) & Lab \\
    $\beta_{\rm sh}$ & Normalised velocity of shock (to $c$) & Lab \\
    $R_{\rm sh}$ & Shock distance from burster & Lab \\
    $\Delta R$ & Fluid shell width & Lab \\
    $B$ & \textbf{B}-field & Lab \\ 
    $\nu$ & Frequency & Observer \\
    $\gamma_{e}$ & Particular Lorentz factor of electron & Fluid \\
    $n_{e}$ & Electron number density (in energy space)  & Fluid \\
    $\tau_{\rm esc}$ & Escape timescale & Fluid \\
    $\tau_{\rm ad}$ & Adiabatic timescale & Fluid \\
    $e$ & Internal energy density & Fluid \\
    $\rho$ & Fluid mass density & Fluid \\
    $\rho_{\rm ext}$ & External mass density & Lab \\
    $n_{\rm ext}$ & External number density & Lab \\
    $\hat{\gamma}$ & Adiabatic index & Fluid \\
    $\sigma_{\rm KN}$ & KN cross-section & Fluid \\
    $\kappa_{\rm SC}$ & Compton Potential & Fluid \\
    $\gamma_{\rm min}/\gamma_{\rm max}$ & Minimum/maximum injection energy & Fluid \\
    $P_{\nu}$ & Emission coefficient & Fluid \\
    $\delta$ & Doppler boosting factor & Lab \\
    $N$ & Number of particles & All \\
    $Q_{e,0}$ & Injection scaling term & All \\
    $Y_{C}$ & Compton Y-parameter at $\nu_{c}$ & All \\
    $\nu_{a}$ & Self-absorption break frequency & Observer \\
    $\nu_{m}$ & Minimum injection break frequency & Observer \\
    $\nu_{c}$ & Cooling break frequency & Observer \\
    $V$ & Volume of emitting region & Fluid \\
    \hline
    \end{tabular} \label{tab:Glossary}
\end{table}

\bsp	
\label{lastpage}
\end{document}